\documentstyle[12pt]{article}

\def\NPB#1#2#3{Nucl. Phys. B{#1} (19#2) #3}
\def\PLB#1#2#3{Phys. Lett. B{#1} (19#2) #3}

\def\PRD#1#2#3{Phys. Rev. D{#1} (19#2) #3}

\def\ts{\textstyle}
\def\yzero{\smash{\hbox{$y\kern-4pt\raise1pt\hbox{${}^\circ$}$}}}

\def\-{\hphantom{-}}
\def\ov{\overline}
\def\s2{\frac{1}{\sqrt2}}
\def\s22{\frac{1}{2\sqrt2}}

\def\wt{\widetilde}
\def\oh{\frac{1}{2}}
\def\beq{\begin{equation}}
\def\eeq{\end{equation}}
\def\beqa{\begin{eqnarray}}
\def\eeqa{\end{eqnarray}}
\def\tr{{\rm tr \,}}

\def\IF{\relax{\rm I\kern-.18em F}}
\def\II{\relax{\rm I\kern-.18em I}}
\def\IP{\relax{\rm I\kern-.18em P}}
\def\IC{\relax\hbox{\kern.25em$\inbar\kern-.3em{\rm C}$}}
\def\IR{\relax{\rm I\kern-.18em R}}

\def\cad{{\cal D}}

\def\cam{{\cal M}}
\def\cn{{\cal N}}

\def\Dsl{\,\raise.15ex\hbox{/}\mkern-13.5mu D} 
\def\IZ{Z\kern-.4em  Z}
\def\cp#1{\relax\ifmmode {\IP\kern-2pt{}_{#1}}\else $\IP\kern-2pt{}_{#1}$\fi}

\begin{document}

\makeatletter
\@addtoreset{equation}{section}
\makeatother
\renewcommand{\theequation}{\thesection.\arabic{equation}}
\pagestyle{empty}
\rightline{ FTUAM-97/6; UCVFC-DF/11-97}
\rightline{\tt hep-th/9706158}
\begin{center}
\LARGE{
 Non-Perturbative Heterotic D=6,4,\  N=1\\[1mm]
Orbifold Vacua \\[5mm]}
\large{G.~Aldazabal$^1$,
A. Font$^2$,
L.~E.~Ib\'a\~nez$^3$,
A.~M.~Uranga$^3$
and G. Violero$^3$\\[2mm]}
\small{
$^1$ CNEA, Centro At\'omico Bariloche,\\[-0.3em]
8400 S.C. de Bariloche, and CONICET, Argentina.\\[1mm]
$^2$ Departamento de F\'{\i}sica, Facultad de Ciencias, 
Universidad Central de Venezuela \\[-0.3em]
A.P. 20513, Caracas 1020-A, Venezuela. \\[-0.3em]
and \\[-0.3em]
Centro de Astrof\'{\i}sica Te\'orica, Facultad de Ciencias,\\[-0.3em]
Universidad de Los Andes, Venezuela.\\[1mm] 
$^3$ Departamento de F\'{\i}sica Te\'orica C-XI
and Instituto de F\'{\i}sica Te\'orica  C-XVI,\\[-0.3em]
Universidad Aut\'onoma de Madrid,
Cantoblanco, 28049 Madrid, Spain. 
\\[4mm]}
\small{\bf Abstract} \\[5mm]
\end{center}

\begin{center}
\begin{minipage}[h]{14.0cm}
We consider $D=6$,  $N=1$, $Z_M$ orbifold compactifications of heterotic
strings in which the usual modular invariance constraints
are violated. It is argued that in the presence of non-perturbative effects  many 
of these vacua are nevertheless consistent. The perturbative massless sector
can be computed explicitly
from the  perturbative mass formula subject  to an extra shift in the
vacuum energy. This shift is associated to a non-trivial
antisymmetric B-field  flux at the orbifold fixed points. 
The non-perturbative piece is given by five-branes either moving 
in the bulk or stuck at the fixed points, giving rise to Coulomb phases 
with tensor multiplets. 
The heterotic duals of  some Type IIB orientifolds belong
 to this class of orbifold models.
We also discuss how to carry out this type of construction to
the $D=4$, $N=1$ case and specific
$Z_M\times Z_M$ examples are presented in which non-perturbative
transitions changing the number of chiral generations do occur.
\end{minipage}
\end{center}
\newpage

\setcounter{page}{1}
\pagestyle{plain}
\renewcommand{\thefootnote}{\arabic{footnote}}
\setcounter{footnote}{0}

\section{Introduction}

While studying $D$-dimensional perturbative heterotic vacua one of the
most important ingredients is one-loop modular invariance. This
property automatically guarantees the absence of gauge and
gravitational anomalies.  This led  some people to believe that
the important  concept  was that of modular invariance in closed string 
theory and that  absence of anomalies was just a mere consequence 
of  that (more important) principle.

With the recent developments of the last three years it has become clear that
closed and open strings are on equal footing and that the fundamental
concept  is  indeed  anomaly cancelation and not modular invariance.  
If that is the case, a natural question arises:  if we consider a
 $D$-dimensional, {\it e.g.} heterotic construction, which does not fullfil the
usual criteria for  one-loop modular invariance,  can it still  
 be consistent  as a quantum theory?  Certainly 
there are examples of  $D=6$ theories whose perturbative sector is
anomalous 
and still are  consistent quantum theories. The prototypes are  heterotic 
$SO(32)$  or $E_8\times E_8$ compactifications  on $K3$ with 
instanton number $k<24$ embedded in the gauge group. In other words,
smooth $K3$ compactifications in which  the size of $24-k$ instantons has been
 set to zero.   These string models have gauge and/or gravitational anomalies
at the perturbative level. However, 
it was shown in refs.~\cite{witsm, dmw}  that there are 
non-perturbative effects (which occur no matter how small the 
string coupling is) which provide additional gauge and/or tensor and/or 
hypermultiplets which render the theories anomaly-free.

It would  be very interesting to know, given a certain
heterotic string compactification which is anomalous at the perturbative 
level,
when it will  be completed by non-perturbative effects that cancel the
anomalies.  In the meantime,  while trying to obtain new
heterotic vacua one should not really impose as a crucial ingredient 
additional constraints associated to modular invariance.

In this paper we apply the above philosophy to the
construction of $N=1$, $D=6,4$  toroidal heterotic  orbifolds.
In the past, perturbative Abelian orbifold vacua  have been extensively studied
\cite{dhvw, orbi}
in which certain modular invariance constraints are imposed on
the  gauge embedding of the $Z_M$ twist.  We consider the modding of
 $SO(32)$ and $E_8\times E_8$  $D=6$ heterotic strings by 
gauge embeddings not verifying  those constraints. 
In order to get consistent spectra certain shifts in the
left-handed vacuum energy (associated to an antisymmetric
B-field flux) must be added. One obtains in this
way  perturbative  string vacua which are naively anomalous. 
We find that 
some of them can be understood as orbifold versions of 
the smooth  $K3$ compactifications with  some instantons set to
zero size \cite{witsm, dmw}. 
Hence, such orbifold models are expected to be 
completed by non-perturbative effects  to produce a final
anomaly-free theory.  They  correspond to the presence of 
 five-branes with their  world-volume in the uncompactified dimensions.

We find that  some orbifold models of the $SO(32)$ 
heterotic string require  the addition of
five-branes stuck at the fixed points.  They seem to correspond  to
theories that are in a Coulomb phase in which tensor multiplets
appear, as in the transitions considered in \cite{intri}. In this class
there are models which seem to be the heterotic duals of the 
$Z_M$, $M=2,3,4,6$,  Type IIB orientifolds of ref.~\cite{gj, dabol1}.
For some other orbifold models in which there are no appropriate twisted 
moduli to repair  the orbifold singularities nor transitions of 
the type studied in ref.~\cite{intri},  we lack at the moment a proper 
interpretation.

One can also consider the construction of $N=1$, $D=4$ orbifold theories with
gauge embeddings not verifying the usual modular invariance constraints. 
However, our knowledge of  the possible non-perturbative effects which could
render such theories consistent is much more incomplete. Still, 
in some $Z_M\times Z_M$, $D=4$, $N=1$ orbifold  constructions,  some 
conclusions  can be extracted  on the basis of $D=6$ information.  In particular,
$D=6$ transitions to a Coulomb phase  in which 29 hypermultiplets  turn into a
tensor multiplet  \cite{sw6d, mv1} 
have as a 4-dimensional consequence transitions in which
the number of chiral ({\it e.g.} $E_6$) generations varies.  A similar 
observation has recently been made in ref.~\cite{kasi}, using a different 
approach.

The organization of this paper goes as follows. In  the next chapter
we discuss some of the known methods to get perturbative 
 $D=6$, $N=1$ heterotic vacua. This  includes $Z_M$,
$(0,4)$ toroidal orbifold constructions  as well as smooth
$K3$ compactifications in the presence of  non-Abelian or 
Abelian gauge bundles.  As a byproduct  we show that, in the same
way that one can characterize $E_8\times E_8$ compactifications by giving 
a pair of instanton numbers $(k_1,k_2)$, one can classify 
$Spin(32)/Z_2$ compactifications in terms of a pair of instanton numbers
$(k_{NA}, k_{A})$ corresponding to a  non-Abelian and an Abelian 
instanton background respectively.
We  provide all the perturbative
$Z_2$ and $Z_3$ symmetric orbifold heterotic  vacua  (without quantized 
Wilson lines) and  discuss  connections  with smooth $K3$ compactifications
as well as with F-theory models. The theories obtained  from smooth 
$K3$ compactifications with zero size instantons are also briefly recalled.

In Chapter 3 we consider the construction  of  $D=6$  
heterotic orbifolds in which the gauge embedding violates the usual 
modular invariance constraints. We construct a
family of $Z_3$ models  that provide orbifold
versions  of $K3$  compactifications  with some zero size
instantons.  They require  the presence of a quantized number of 
five-branes in order to cancel anomalies. The different 
behaviour of small  instantons in $SO(32)$ and $E_8\times E_8$  
naturally appears.  In the case of the standard embedding, there exist
possible transitions in which  2 instantons per fixed point are 
converted into  2 five-branes.  
Some properties of this
class of models are further clarified if one considers 
the relevant index theorem formulae for instantons on
$Z_M$ ALE spaces.  This  is done in Chapter 4.

In Chapter 5  we describe non-perturbative $Z_3$ orbifold models 
on $SO(32)$ which seem to correspond to vacua in  a
Coulomb phase in which tensor multiplets appear. These are
the first explicit  heterotic compactifications realizing the
transitions considered in ref.~\cite{intri}  which appear when a certain
minimum number of  small instantons  accumulate on a $Z_M$ 
singularity.  The simplest  example of these  models corresponds to
the heterotic dual of the $Z_3^A$ Type IIB orientifold of
ref.~\cite{gj, dabol1}.
In Chapter 6 we discuss some aspects of even $M$, $Z_M$ orbifolds.
In particular, we show how the heterotic duals  of the $Z_4^A$ and
$Z_6^A$ Type IIB orientifolds  of ref.~\cite{gj, dabol1} 
can be understood as  $SO(32)$ 
heterotic orbifolds with certain (non-modular invariant)  gauge embeddings.
The same is true for some other  $Z_2$  orientifolds 
considered  in ref.~\cite{dabol2, gm}.
All these  theories are in a Coulomb phase so that  extra tensor multiplets 
appear. We also show how  the  $Z_2$ orientifold of ref.~\cite{bso, gp}   
(which we will call from now on the BSGP model) 
with all five-branes at the same fixed point  seems to be  dual to  a $Z_2$ heterotic  
orbifold with a non-modular invariant shift. Other examples of  $Z_4$ 
orbifolds but with 
vector structure are also presented.

In Chapter 7  we discuss the possibility of extending this kind of analysis to
$D=4$, $N=1$ models. As mentioned before,  
in certain $Z_N\times Z_M$  orbifolds one can obtain some partial
information in terms of  the  $D=6$ subsectors of  the theory. 
The $Z_2\times Z_2 $, $D=4$  Type IIB orientifold of ref.~\cite{bl}  
can be understood 
as a certain $Z_2\times Z_2$ orbifold of the SO(32) heterotic string in which the
$Z_2$ gauge embeddings  do not verify the usual modular invariance constraints.
We consider analogous $Z_3\times Z_3$ heterotic orbifolds and show how 
$D=6$ transitions to Coulomb phases  reflect themselves in four dimensions 
into transitions that can change  the number of chiral generations. 
As an example, the standard $Z_3\times Z_3$,
$E_8\times E_8$  orbifold with discrete
torsion appears to have non-perturbative transitions to a model with just
three $E_6$ generations from the untwisted sector.
Some final comments are left for Chapter 8.

\section{Perturbative Heterotic D=6, N=1 Vacua}

Before introducing the non-perturbative 
heterotic orbifold models let us recall some facts about
perturbative $D=6$, $N=1$ vacua 
which will be of interest
for the later discussion. 
An interesting class of $D=6$, $N=1$ heterotic vacua can be obtained
from  symmetric toroidal orbifold compactifications on $T^4/Z_M$.
The construction of these models parallels that of $T^6/Z_M$ orbifolds
\cite{dhvw,orbi} as considered in refs.~\cite{walton, erler,afiq}. Here we
briefly review the notation and the salient points relevant to our discussion. 
Acting on the (complex) bosonic transverse coordinates,
the $Z_M$ twist $\theta$ has eigenvalues $e^{2 \pi i\, v_a}$,
where $v_a$ are the components of $v=(0,0,\frac1M,-\frac1M)$.
$M$ can take the values $M=2,3,4,6$. The embedding
of $\theta$ on the gauge degrees of freedom is usually
realized by a shift $V$ such that $MV$ belongs to the $E_8\times E_8$ lattice 
$\Gamma_8 \times \Gamma_8$ or to the  $Spin(32)/Z_2$ lattice $\Gamma_{16}$.
This shift is restricted by the modular invariance constraint 
\beq
M\, (V^2-v^2)={\rm even}
\label{dos}
\eeq
All possible embeddings for each $M$ are easily found.
In the $E_8\times E_8$ case, we find 2 inequivalent embeddings
for $Z_2$, 5 for $Z_3$, 12 for $Z_4$ and 59 for $Z_6$,
leading to different patterns of $E_8\times E_8$ 
breaking to rank 16 subgroups.
For $Spin(32)/Z_2$ we find 3 inequivalent embeddings
for $Z_2$, 5 for $Z_3$, 14 for $Z_4$ and 50 for $Z_6$.
Each of these models is only the starting point
of a bigger class of vacua, generated by adding Wilson
lines in the form of further shifts in the gauge
lattice satisfying extra modular invariance constraints,
by permutations of gauge factors, etc..

The spectrum for each model is subdivided in sectors. There are $M$ sectors 
twisted by $\theta^j, j=0,1,\cdots , M-1$. Each particle state is created by 
a product of left and right vertex operators  $L\otimes R$. At a generic point 
in the four-torus moduli space, the massless states follow from
\beq
\label{uno}
m_R^2=N_R+\frac{1}{2}\, (r+j\,v)^2
+E_n-\frac{1}{2} \quad ;\quad
m_L^2=N_L+\frac{1}{2}\, (P+j\,V)^2+E_j-1
\eeq
Here $r$ is an $SO(8)$ weight with $\sum_{i=1}^4 r_i=\rm odd$
and $P$ a  gauge lattice vector with $\sum_{I=1}^{16} P^I= \rm even$.
$E_j$ is the twisted oscillator contribution to the zero point energy and it 
is given by $E_j=j(M-j)/M^2$. The multiplicity of states satisfying 
eq.~(\ref{uno}) in a $\theta^j$ sector is given by the appropriate 
generalized GSO projections \cite{afiq}. In the untwisted sector there
appear the gravity multiplet, a tensor multiplet, charged hypermultiplets
and 2 neutral hypermultiplets (4 in the case of $Z_2$). In the twisted
sectors only charged hypermultiplets appear.
The generalized GSO projections are particularly simple in the $Z_2$
and $Z_3$ case since essentially all massless states survive with the same
multiplicity. 
The spectra for all $Z_2$ and 
$Z_3$ embeddings are shown in Tables~\ref{tabla1} and \ref{tabla2}.

\begin{table}[p]
\footnotesize
\renewcommand{\arraystretch}{1.25}
\begin{center}
\begin{tabular}{|c|c|c|c|}
\hline
Shift $V$ & & & \\
\cline{1-1}
Group & \raisebox{2.5ex}[0cm][0cm]{Untwisted matter} &
\raisebox{2.5ex}[0cm][0cm]{Twisted matter} &
\raisebox{2.5ex}[0cm][0cm]{$(k_1,k_2)$} \\
\hline\hline
$\frac12(1,1,0, \cdots, 0)\times (0, \cdots, 0)$ &
(56,2)+4(1,1) &
8(56,1)+32(1,2)$^*$  & (24,0) \\
\cline{1-1}
$E_7\times SU(2)\times E_8$ & & & \\
\hline\hline
$\frac12(1,0, \cdots, 0)\times (1,1,0 \cdots, 0)$ &
 (1,56,2)+4(1,1,1) &
8(16,1,2) & (16,8) \\
\cline{1-1}
$SO(16)\times E_7\times SU(2)$ & + (128,1,1) &  & \\
\hline
\multicolumn{4}{c}{} \\[-0.5cm]
\hline
$\frac13(1,1,0, \cdots, 0)\times (0, \cdots, 0)$ &
(56,1)+3(1,1) &
9(56,1)+18(1,1)$^*$  & (24,0) \\
\cline{1-1}
$E_7\times U(1)\times E_8$ & & + 45 (1,1)$^*$ & \\
\hline\hline
$\frac13(2,0, \cdots, 0)\times \frac13(2,0 \cdots, 0)$ &
(14,1)+(64,1) + &
9(14,1)+9(1,14)  & (12,12) \\
\cline{1-1}
$SO(14)\times SO(14)\times U(1)^2$ & (1,14)+(1,64) + 2(1,1) & + 18(1,1)$^*$ & \\
\hline\hline
$\frac13(1,1,1,1,2,0,0,0)\times (0, \cdots, 0)$ &
(84,1)+2(1,1) &
9(36,1)+18(9,1)$^*$  & (24,0) \\
\cline{1-1}
$SU(9)\times E_8$ & & & \\
\hline\hline
$\frac13(1,1,2,0, \cdots, 0)\times \frac13(1,1,0 \cdots, 0)$ &
(27,3,1) + (1,1,56)   &
9(27,1,1)+9(1,3,1)  & (18,6) \\
\cline{1-1}
$E_6\times SU(3)\times E_7\times U(1)$ & + 3(1,1,1) & + 18(1,3,1)$^*$ & \\
\hline\hline
$\frac13(1,1,1,1,2,0 \cdots, 0)\times \frac13(1,1,2,0,0,0)$ &
(1,27,3) + (84,1,1)   &
9(9,1,3) & (15,9) \\
\cline{1-1}
$SU(9)\times E_6\times SU(3)$ & + 2(1,1,1) & & \\
\hline
\end{tabular}
\end{center}
\caption{Perturbative $Z_2$ and $Z_3$, $E_8\times E_8$, orbifold models. The asterisk 
indicates twisted states involving left-handed oscillators. The last column 
shows which smooth $K3$ compactification yields a similar massless spectrum 
{\it after Higgsing}.
\label{tabla1} }
\end{table}

\begin{table}[p]
\footnotesize
\renewcommand{\arraystretch}{1.25}
\begin{center}
\begin{tabular}{|c|c|c|c|}
\hline
Shift $V$ & & & \\
\cline{1-1}
Group & \raisebox{2.5ex}[0cm][0cm]{Untwisted matter} &
\raisebox{2.5ex}[0cm][0cm]{Twisted matter} &
\raisebox{2.5ex}[0cm][0cm]{$G_0$} \\
\hline\hline
$\frac12(1,1,0, \cdots, 0)$ &
(28,2,2)+4(1,1,1) &
8(28,1,2)+32(1,2,1)$^*$  & $SO(8)$ \\
\cline{1-1}
$SO(28)\times SU(2)\times SU(2)$ & &  & \\
\hline\hline
$\frac12(1,1,1,1,1,1,0, \cdots, 0)$ &
(12,20)+4(1,1)  &
8(32,1) & $SO(8)$ \\
\cline{1-1}
$SO(12)\times SO(20)$ & &  &  \\
\hline\hline
$\frac14(1, \cdots, 1,-3)$ &
(120) + ($\ov{120}$)  &
8(16) + 8($\ov{16}$) & $\II$ \\
\cline{1-1}
$SU(16)\times U(1)$ & + 4(1) &  &  \\
\hline
\multicolumn{4}{c}{} \\[-0.5cm]
\hline
$\frac13(1,1,0, \cdots, 0)$ &
(28,2)+3(1,1) &
9(28,2)+18(1,1)$^*$  & $SO(8)$ \\
\cline{1-1}
$SO(28)\times SU(2)\times U(1)$ & & + 45 (1,1)$^*$ & \\
\hline\hline
$\frac13(1,1,1,1,2,0, \cdots, 0)$ &
(22,5)+(1,10)  &
9(22,1)+9(1,10)  & $SO(8)$ \\
\cline{1-1}
$SO(22)\times SU(5)\times U(1)$ & + 2(1,1) & + 18(1,5)$^*$ &  \\
\hline\hline
$\frac13(1,1,1,1,1,1,1,1,0,\cdots, 0)$ &
(16,8)+(1,28) &
9(1,28)+18(1,1)$^*$  &  $SO(8)$\\
\cline{1-1}
$SO(16)\times SU(8) \times U(1)$ & + 2(1,1) & & \\
\hline\hline
$\frac13(1,\cdots,1,2,0,0,0,0,0)$ &
(10,11) + (1,55)   &
9(1,11)+9(16,1)  & $\II $ \\  
\cline{1-1}
$SO(10)\times SU(11) \times U(1)$ & + 2(1,1) & & \\
\hline\hline
$\frac13(1,\cdots , 1,0,0)$ &
(14,2,2) + (91,1,1)   &
9(1,1,1) + 9(14,2,1) & $\II $ \\  
\cline{1-1}
$SU(14)\times U(1) \times SU(2)\times SU(2)$ &  + 2(1,1,1) & + 
18(1,1,2)$^*$& \\ 
\hline
\end{tabular}
\end{center}
\caption{Perturbative $Z_2$ and $Z_3$, $Spin(32)/Z_2$, orbifold models. The asterisk 
indicates twisted states involving left-handed oscillators. The last column 
shows the generic terminal gauge group $G_0$ after Higgsing.
\label{tabla2} }
\end{table}

It is instructive to compare these orbifold vacua with the $D=6$, $N=1$ 
models obtained upon generic heterotic compactifications on 
smooth $K3$ surfaces in the presence of instanton backgrounds
\cite{kv, dmw, sw6d}.
In the $E_8\times E_8$ case there are instanton numbers $(k_1,k_2)$  
satisfying $k_1+k_2=24$, as required by anomaly cancelation.
It is convenient to define $k_1=12+n $, $ k_2= 12-n $
and assume $n\ge 0$ without loss of generality. For $n \leq 8$,
an $SU(2)$ background on each $E_8$ leads to $E_7\times E_7$ 
unbroken gauge group with hypermultiplet content 
\beq
\oh(8+n)({\bf 56},{\bf 1}) + \oh (8-n)({\bf 1}, {\bf 56})
+  62({\bf 1}, {\bf 1})
\label{e7e7}
\eeq
Due to the pseudoreal character of the ${\bf 56}$ of $E_7$,
odd values of $n$ can also be considered.

Understanding the spectrum corresponding to $n>8$ requires some knowledge 
on $E_8\times E_8$ vacua in the presence of five-branes. In 
non-perturbative $E_8\times E_8$ compactifications on $K3$, 
cancelation of gravitational anomalies requires
\beq
k_1+k_2+n_B=24
\label{anombe8}
\eeq
where $n_B$ is the number of $E_8\times E_8$ five-branes. These are
five-branes of M-theory, each one carrying a tensor multiplet and a
hypermultiplet \cite{dmw}.
When $9 \leq n \leq 12$, $k_2$ is not large enough to support an  
$SU(2)$ background, the instantons become small and turn into
five-branes that give rise to extra tensor multiplets. The unbroken gauge 
group is now $E_7 \times E_8$ with hypermultiplets  \cite{sw6d}
\beq
\oh(8+n)({\bf 56},{\bf 1}) + (53+n)({\bf 1}, {\bf 1})
\label{e7e8}
\eeq
and $(12-n)$ extra tensor multiplets.

Models with various groups can be obtained from (\ref{e7e7})
and (\ref{e7e8}) by symmetry breaking. The group from the second $E_8$
does not possess, in general, enough charged matter to be completely broken.
Higgsing stops at some terminal group, depending on the value of $n$, 
with minimal or no charged matter \cite{kv, afiq}. 
For instance $E_8$, $E_7$, $E_6$, $SO(8)$, $SU(3)$ 
terminal groups are obtained for $n=12,8,6,4,3$ while complete breaking 
proceeds for $n=2,1,0$. On the other hand, the first $E_7$ can be completely 
Higgsed away.  In the last column of Table~\ref{tabla1} we show the instanton numbers
$(k_1,k_2)$ of compactifications yielding, upon Higgsing, 
a massless spectrum similar to the corresponding orbifold. We thus see that the
five $Z_3$ orbifolds of $E_8\times E_8$  are in the same
moduli space as generic $K3$ compactifications with $n=12,0,12,6,3$ 
respectively. The two $Z_2$ orbifolds correspond to $n=12,4$ respectively.
This connection between modular invariant orbifold models and
instanton backgrounds will be further clarified in Chapter 4.

In the $Spin(32)/Z_2$ case, embedding a total instanton number $k=24$ is required
to cancel gravitational anomalies. An $SU(2)$ background breaks the symmetry down 
to $SO(28)\times SU(2)$ with hypermultiplets in 
$10({\bf 28},{\bf 2})+65({\bf 1},{\bf 1})$.  Hence,
upon Higgsing, the generic group is $SO(8)$.  This class
of models is known to be  \cite{mv1, berkooz}
 in the same moduli space as $(k_1,k_2)=(16,8)$ 
compactifications of $E_8\times E_8$. As shown in Table~\ref{tabla2}, 
the first three  $Z_3$ orbifolds  of $Spin(32)/Z_2$  do have $SO(8)$ as 
generic group 
but the last two models have trivial gauge group after full Higgsing.

In fact this is not completely new, it was already noticed in 
\cite{afiu} that the fourth $Spin(32)/Z_2$, $Z_3$ model, could 
lead to complete Higgsing. Also,
in ref.~\cite{berkooz} the authors construct a heterotic $Z_2$ orbifold,
`without vector structure', in which the resulting $U(16)$ group can
be completely broken. In our language this $Z_2$ orbifold has embedding
$V=\frac14(1,\cdots,1,-3)$
 (third example in Table~\ref{tabla2}).
In general, embeddings with vector structure have shifts $V$ such that
$MV=(n_1, \cdots , n_{16})$, whereas embeddings without vector structure 
have $MV=(n_1 + \oh, \cdots , n_{16} +\oh)$. Since $MV \in \Gamma_{16}$,
$\sum_I n_I = {\rm even}$ in both cases.

Beyond perturbation theory the condition $k=24$ is replaced by 
\beq
k+n_B=24
\label{anombso} 
\eeq
where $n_B$ is now the number of dynamical $SO(32)$ five-branes, which can be
understood as small instantons \cite{witsm}. One such brane carries an $Sp(1)$
vector multiplet, but when $r$ of them coincide at a point on the smooth
$K3$, the group is enhanced to $Sp(r)$. In general, the non-perturbative
group is $\prod Sp(r_i)$ with $\sum r_i = n_B$. The five-branes also carry
non-perturbative hypermultiplets. In particular, for each $Sp(r)$ there
appear 32 half hypermultiplets in the fundamental representation, together
with one hypermultiplet in the antisymmetric two-index representation
(decomposable as a singlet plus the rest). Cancelation of gauge anomalies
requires that the hypermultiplets in the fundamental representation be also
charged under the perturbative gauge group that arises when $SO(32)$ is
broken by the background with instanton number $k=24-n_B$. 
The results just summarized apply to compactifications on smooth $K3$ 
surfaces. 
When the $K3$ is realized as an orbifold, the five-branes can coincide at a
fixed point thereby producing other gauge groups and hypermultiplet content
(see  below).

We have seen that the   $E_8\times E_8$ compactifications can be labeled by
a pair of instanton numbers $(k_1,k_2)$ with
$k_1=12+n$, $k_2=12-n$ and $n=0,\cdots ,12$.
Recently it has become clear that there are in fact different types of
$Spin(32)/Z_2$ instantons depending on the generalized second 
Stieffel-Whitney class \cite{berkooz}.
An analysis in terms of F-theory  \cite{aspfz2} has shown that in a 
general $Spin(32)/Z_2$ heterotic compactification, instantons with and without 
vector structure are present, their contribution to the total instanton 
number being respectively $8+4n$ and $16-4n$, with the integer $n$ 
satisfying $-2\leq n \leq 4$. 
A simple heterotic realization of this idea can be 
obtained by embedding a $U(1)\times SU(2)$ background in 
$SO(32) \supset  SU(16) \times U(1) \supset
SU(14)\times U(1)^{\prime} \times U(1) \times SU(2)$.
Then the $Spin(32)/Z_2$ vacua can be labeled by 
giving the pair of instanton numbers $(k_{NA}, k_A)$ 
with $k_{NA}=8+4n$ and $k_A=16-4n$.
The adjoint decomposition is
\beqa
{\bf 496} & = & 
({\bf 1},0,0,{\bf 3}) + ({\ov {\bf 14}},\oh,0,{\bf 2}) + 
({\bf 14},-\oh,0,{\bf 2}) + ({\bf 195},0,0,{\bf 1}) + 2({\bf 1},0,0,{\bf 1})
+ \nonumber  \\
& {} & ({\bf 1},1,\s22,{\bf 1}) + ({\bf 14},\oh,\s22,{\bf 2}) + 
({\bf 91},0,\s22,{\bf 1}) 
+  \nonumber \\
& {} & ({\bf 1},-1,-\s22,{\bf 1}) + ({\ov{\bf 14}},-\oh,-\s22,{\bf 2}) + 
({\ov{\bf 91}},0,-\s22,{\bf 1}) 
\label{auno}
\eeqa
where the two middle entries denote the $U(1)^{\prime}\times U(1)$ charges.
The massless spectrum that arises upon embedding $k_A=(16-4n)$ instantons in $U(1)$
and $k_{NA}=(8+4n)$ in $SU(2)$ is found
using the index theorem formulae \cite{gsw,afiu}. For
$-1\leq n\leq 2$ we find the following $SU(14)\times U(1)^{\prime} \times U(1)$ 
hypermultiplets
\beqa
& {} & (1-\frac n2)({\bf 1},1,\s22) + (1-\frac n2)({\bf 1},-1,-\s22) + 
(1-\frac n2) ({\bf 91},0,\s22) + \nonumber \\
& {} & (1-\frac n2)({\ov{\bf 91}},0,-\s22) + (6+n)({\ov{\bf 14}},-\oh,-\s22) + 
(6+n)({\bf 14},\oh,\s22) + \nonumber \\
& {} &(2+2n)({\bf 14},-\oh,0) + (2+2n)({\ov{\bf 14}},\oh,0) 
+ (33+8n)({\bf 1},0,0)
\label{ados}
\eeqa

For $n=3$ there are not enough instantons to support the $U(1)$ bundle. 
The corresponding instantons become small and give the spectrum of a 
pointlike instanton without vector structure \cite{aspfz2}. The resulting 
model has a gauge group $SO(28)\times SU(2)\times 
Sp(4)$, a hypermultiplet content
\beq
8({\bf 28}, {\bf 2},{\bf 1}) + 56({\bf 1},{\bf 1},{\bf 1}) + \oh({\bf 
28},{\bf 1},{\bf 8}) + ({\bf 1},{\bf 2},{\bf 8})
\label{atres}
\eeq
and one additional tensor multiplet.
For $n=4$  instantons without vector structure 
 disappear and one just has the $SU(2)$ bundle 
with 24 instantons mentioned above. For $n=-2$, the situation is reversed, 
since there is left just a
$U(1)$ bundle with 24 instantons. The resulting  gauge group is $U(16)$ with
hypermultiplets
\beq
2({\bf 120}, \frac1{2\sqrt2}) + 2({\bf \ov{120}}, -\frac1{2\sqrt2})
+ 20({\bf 1},0)
\label{acuatro}
\eeq
For each value of $n$, appropriate sequential Higgsing produces chains of 
models that match similar $E_8\times E_8$ heterotic chains \cite{afiq}, 
for the same value of $n$, thus providing several identifications between 
compactifications of both heterotic strings. This equivalence is evident 
in the F-theory framework, since the Calabi-Yau spaces obtained upon 
Higgsing (taking generic polynomials in the fibration over $\IF_n$) are 
identical in both types of chains. Also, on the heterotic side 
perturbative T-dualities have been shown to relate $Spin(32)/Z2$ and 
$E_8\times E_8$ compactifications for several values of $n$ \cite{berkooz}. 
By Higgsing it can be shown that the $Z_3$  
models listed in Table~\ref{tabla2} correspond to 
$n=4,4,4,1,1$, respectively. The three $Z_2$ models correspond to 
$n=4,4,0$.

To summarize,  we see that standard orbifold 
compactifications provide a rich variety of perturbative $D=6$ vacua, and 
have an interesting interplay with other techniques. Our purpose in this 
article is to extend this to non-perturbative constructions.

\section{Non-Perturbative $D\!=\!6$ Heterotic Orbifolds and Small Instantons}

\subsection{Non-Modular Invariant Heterotic Orbifolds}
\label{nmihm}

As reviewed in the previous section,
different types of non-perturbative phenomena occur when
the size of the instantons on  a (smooth) $K3$ compactification goes to zero.
The   instanton singularity is softened by the presence of new 
multiplets of non-perturbative origin. This yields  new classes
of consistent  $D=6$, $N=1$  non-perturbative heterotic vacua.

In this type of theories there is a perturbative part corresponding 
to the  $E_8\times E_8$ or $Spin(32)/Z_2$ degrees of freedom with a 
background  
of (large) instantons  with total instanton number {\it smaller } than 24.
There is also a non-perturbative piece which is not
describable by usual perturbative string theory techniques.  
Although the perturbative side of these models is well understood, it has the
shortcoming that generically this kind of $K3$ compactifications are not
free conformal field theories (CFT).  One would like  to construct 
analogous heterotic vacua  in which the  perturbative part 
is just some free CFT like, in particular, a toroidal $Z_M$ orbifold.
Furthermore,  $D=6$ orbifolds of M-theory \cite{dasmu, gm}, 
F-theory \cite{gm, bz, blum, senft}, 
and IIB  orientifolds  \cite{bso, hor, bs} 
have been recently used  \cite{gj, gp, bz, dabol1, dabol2, bianchi} 
to construct new  vacua. It
would be interesting to see whether there are connections between those 
models and orbifolds of heterotic theory. 

Thus, as a first  step, we would like to construct  
in this chapter  non-perturbative vacua
in which the perturbative side is describable by some sort of  standard
heterotic $Z_M$ orbifold.  It is clear that this perturbative piece need not be
modular invariant by itself, the crucial criterium would be now  
anomaly cancelation in the {\it complete } theory.  Global 
consistency of the theory requires the 
fulfillment of the constraint
\beq
\int  _{X} dH\  =\ 0
\label{iuno}
\eeq
\label{anomper}
where $H$ is the three-form heterotic field strength
and  $X$ is the compact space (orbifold).  In the presence of five-branes and/or
orbifold fixed points in the  $X$  manifold,  there are  extra   sources  for $ dH$.
Already at the perturbative level, there can be  a non-vanishing flux of
the antisymmetric  tensor at  orbifold singularities.  At the
non-perturbative level,  there can be five-branes   whose  worldvolume
fills the  6  uncompactified dimensions  and which act as magnetic sources
for the antisymmetric field.  They can be understood as duals of type I 
D-branes 
in the $SO(32)$ case and as M-theory  five-branes in the $E_8\times E_8$  
case. In both cases the total charge associated to the antisymmetric
tensor must vanish. This is
\beq
\sum _f \ Q_f \ + \  n_B \ =\ 0
\label{idos}
\eeq
\label{conserv}
where $Q_f$ is the magnetic charge associated to each fixed point  
and $n_B$ is the number of five-branes present.  
We would like first  to obtain  heterotic orbifold models that would be 
the analogue of  the smooth  $K3$ compactifications in the
presence of  five-branes. 
In constructing this class of non-perturbative orbifold models we will  be guided 
by the {\it smooth}  $K3$ compactifications of $E_8\times E_8$  in the
presence of wandering five-branes  considered in refs.~\cite{dmw, sw6d}.
In order to make contact with these   smooth compactifications we have to consider
$Z_M$ orbifolds in which there are massless oscillator modes appropriate to  repair the
orbifold  singularities.   

Let us start with the perturbative sector of the orbifold. 
We can represent the action on the $E_8\times E_8$ or
$SO(32)$   degrees of freedom by a shift $V$ in the gauge lattice.
However, in searching  for consistent models, we are now allowed 
to give up  the modular invariance constraint eq.(\ref{dos}) since
 there are extra contributions which can help to cancel anomalies.
We are interested in computing the massless spectra of such kind of models.
The untwisted perturbative sector is obtained as in usual orbifold models 
and contains the $N=1$, $D=6$ supergravity sector, gauge multiplets  in 
the $E_8\times E_8$  (or $SO(32)$) 
 subgroup invariant under the shift $V$, charged hypermultiplets
and  the untwisted singlet  moduli hypermultiplets.
However the twisted sectors are novel in some respects since the
fixed points are  now sources for the 
 antisymmetric field. Since the right-handed sector
is supersymmetric, we do not expect any special change for the corresponding
zero modes of the mass formula.  However, for the left-moving string it is
reasonable to expect the presence of an appropriate shift in the vacuum level
due to the antisymmetric field flux.
Thus we will  allow for an extra term  $E_B(j)$ in the corresponding
$\theta ^j$  twisted sector  mass formula
\beq
\label{masaI}
m_L^2\ =\ N_L+\frac{1}{2}\, (P+j\,V)^2\ +\ E_j \ +\ E_B(j) \  -1
\eeq
Matching between left and right-movers will thus require
\beq
\label{modmi}
M\, (V^2-v^2)\ +\  2ME_B(j)\ =\ {\rm even}
\eeq
for an order $M$ twisted sector.  This can be understood as an
orbifold version of the Bianchi identity
$dH=\tr F^2- \tr R^2$  as  further discussed  in the next chapter.
If we  are interested in vacua connected to smooth $K3$ compactifications,
we have to  impose the existence of oscillator moduli associated to the 
blowing-up of the singularities at the fixed points.  For $M=2$  one only finds 
$N_L=\frac12$ solutions  of eq.(\ref{masaI})  for $V=0$.  The  $M=3$ case
is richer  and  we discuss it in detail in the following,  postponing the case 
of even $N$ to Chapter 6.

\subsection{A Class of Non-Perturbative Heterotic $Z_3$ Orbifolds}

For $Z_3$ orbifolds we only need consider the $\theta$ sector with
extra energy shift $E_B$.
The two moduli associated to each of the nine fixed points of the orbifold
are given by $\alpha^i_{-\frac13} |0 \rangle$, where $i=1,2$, label the 
two compact complex dimensions and $|0 \rangle$ is a twisted
ground state, singlet under the non-Abelian gauge group of the model.
Such $N_L=\frac13$ oscillator modes will only be massless for
\beq
\label{lengths}
V^2\ =\   \frac89 \ -\  2\ E_B
\eeq
Thus the maximum shift in the vacuum energy will correspond to 
$E_B=\frac49$  (obtained for  $V=0$).  The other extreme case is $V^2=\frac89$,
in which we have $E_B=0$   corresponding  to  some of the modular 
invariant (perturbative) models displayed in Tables~\ref{tabla1} and \ref{tabla2}.
Models with $V^2\ > \frac89$ will not have  the required  oscillator states
in their twisted spectra.  (We will see in the next chapters that  these
longer shifts  are of interest in certain $SO(32)$ heterotic transitions
involving tensor multiplets, but we disregard them for the time being.)

Let us consider first the $SO(32)$ heterotic string.  Consider the class
of shifts  $V$  with  $3V\in \Gamma _{16} $  of the form
\beq
\label{shifts}
V\ =\ \frac13 (1,\cdots , 1,0, \cdots ,0)
\eeq
with an even number $m$  of  $\frac13$ entries
and $m\leq 8$. The unbroken group is $U(m)\times SO(32-2m)$  
and the untwisted sector contains hypermultiplets transforming as
$({\bf  m}, {\bf   32-2m})$ $+({\bf   \frac{m(m-1)}2}, {\bf 1})$ $+2({\bf 1},{\bf 1})$.
The twisted sectors have an extra vacuum shift $E_B= \frac{(8-m)}{18}$ and the mass 
formula  gives massless hypermultiplets in each twisted sector transforming  as
\beq
\label{tspec}
({\bf   \frac{m(m-1)}2}, {\bf 1}) \ +\  2({\bf 1},{\bf 1})
\eeq
for $m=0,2,4,6,8$.  
 
There are two other $Z_3$ models with
singlet  moduli in the twisted sector.  One of them, with
shift $V=(\frac23,0, \cdots,0)$,  has gauge group
 $SO(30)\times U(1)$ and its spectrum is given in Table~\ref{tabla3}. 
The other model has shift $V=\frac16(1, \cdots ,1)$,  $3V$ being a spinorial
weight.  It is thus a $SO(32)$ embedding without vector structure,
a $Z_3$ analogue of  the $Z_2$  orientifolds constructed  in \cite{bso, gp}.
The gauge group is $U(16)$  and the hypermultiplets are also displayed in 
Table~\ref{tabla3}, along with the spectra of the rest of these models.

\begin{table}[p]
\footnotesize
\renewcommand{\arraystretch}{1.25}
\begin{center}
\begin{tabular}{|c|c|c|c|c|}
\hline
Shift $V$ & & & $E_B$ & \\
\cline{1-1}\cline{4-4}
Group & \raisebox{2.5ex}[0cm][0cm]{Untwisted matter} &
\raisebox{2.5ex}[0cm][0cm]{Twisted matter} & $n_B$ &
\raisebox{2.5ex}[0cm][0cm]{$k$} \\
\hline\hline
$(0, \cdots, 0)$ &
2(1,1) &
18(1,1)$^*$ & $\frac49$ & 0 \\
\cline{1-1}\cline{4-4}
$SO(32)$ & & & 24 & \\
\hline\hline
$\frac13(1,1,0, \cdots, 0)$ &
(28,2)+3(1,1)  &
9(1,1)+ 18(1,1)$^*$  & $\frac39$ & 6 \\
\cline{1-1}\cline{4-4}
$SO(28)\times SU(2)\times U(1)$ &  &  & 18 & \\
\hline\hline
$\frac13(2, \cdots,0)$ &
(30) &
9(30)+18(1)$^*$  & $\frac29$ & 12 \\
\cline{1-1}\cline{4-4}
$SO(30)\times U(1) $ & + 2(1,1)& & 12 & \\
\hline\hline
$\frac13(1,1,1,1,0, \cdots, 0)$ &
(24,4) + (1,6)   &
9(1,6)+18(1,1)$^*$  & $\frac29$ & 12 \\
\cline{1-1}\cline{4-4}
$SO(24)\times SU(4) \times U(1)$ & + 2(1,1) &  & 12 & \\
\hline\hline
$\frac13(1,1,1,1,1,1,0, \cdots, 0)$ &
(20,6) + (1,15)   &
9(1,15) + 18(1,1)$^*$  & $\frac19$ & 18 \\
\cline{1-1}\cline{4-4}
$SO(20)\times SU(6)\times U(1)$ & + 2(1,1) &  & 6 & \\
\hline\hline
$\frac16(1, \cdots,1)$ &
(120)+2(1)  &
18(1)$^*$ & $\frac29$ & 12 \\
\cline{1-1}\cline{4-4}
$SU(16)\times U(1)$ & & & 12 & \\
\hline\hline
$\frac13(1,1,1,1,1,1,1,1,0, \cdots,0)$ &
(16,8)+(1,28)  &
9(1,28)+18(1,1)$^* $ & $0 $ &  24  \\
\cline{1-1}\cline{4-4}
$SO(16)\times SU(8) \times U(1)$ &  + 2(1,1)& &  $0$  & \\
\hline\hline
$\frac13(1,1,1,1,1,1,1,1,0, \cdots,0)$ &
(16,8)+(1,28)  &
9(1,1) & $\frac39$ & -- \\
\cline{1-1}\cline{4-4}
$SO(16)\times SU(8) \times U(1)$ &  + 2(1,1)& & 9 T & \\
\hline
\end{tabular}
\end{center}
\caption{Non-Perturbative $Z_3$, $Spin(32)/Z_2$, orbifold models. 
$n_B$ gives the number of five-branes needed to obtain cancelation 
of anomalies. The last example, discussed in Chapter 5, 
requires instead just nine tensor multiplets and no enhanced gauge group
nor extra hypermultiplets \label{tabla3}. It is connected  with the 
next to last model which is in fact perturbative. }
\end{table}

Except for the $m=8$ case, the rest of these models, as they stand, 
have gauge and gravitational anomalies  and the  corresponding 
shifts do not fulfill the perturbative modular invariance constraints.  
However, it turns out  that the addition of an appropriate number of
five-branes renders them consistent, very much in the same way that
a smooth  $K3$ compactification with instanton number $k\ < 24$ becomes 
consistent upon including $24-k$ small instantons (five-branes).
Indeed, one can check that adding $3(8-m)$  five-branes to the 
vacua in eq.(\ref{shifts}) (12 five-branes in the other two cases) leads to 
anomaly-free results. If we consider all of the five-branes coinciding at the same
point (and away from singularities) a non-perturbative gauge  group $Sp(n_B)$ 
is expected to appear, along with hypermultiplets transforming as
\beqa
& {} & \frac12({\bf m},{\bf 1}, {\bf 2n_B})+ \frac12({\bf \ov{m}},{\bf 1}, {\bf 2n_B}) 
+ \frac12({\bf 1},{\bf 32-2m},{\bf 2n_B}) \nonumber \\
& + &
({\bf 1},{\bf 1},{\bf \frac{2n_B(2n_B-1)}2 -1}) + ({\bf 1},{\bf 1},{\bf 1})
\label{bcomple}
\eeqa
with respect to $U(m)\times SO(32-2m)\times Sp(n_B)$.
It is straightforward to check that all 
non-Abelian gauge and gravitational
anomalies do cancel.  Thus, our construction provides a new class of 
consistent  non-perturbative orbifold heterotic vacua.

Notice that the models obtained require the addition of  $6s$, $s=4,3,2,1,0$,  
five-branes.  They contribute one unit of magnetic charge each. Thus, in order
to have an overall vanishing magnetic charge, each of the fixed points
(which in these  particular models are identical) must carry magnetic charge
$Q_f=-\ts{\frac{n_B}{n_f}}$, $n_f$ being the number of fixed points
(9 in the case at hand).
Thus the fixed points carry charges $Q_f=-\frac83,-2,-\frac43,$
$-\frac43,-\frac23,-\frac43$
respectively for each of the first six models in Table~\ref{tabla3}.  Notice 
also 
that the shift in the left-handed vacuum energy is in each case given
by $E_B=-\frac{Q_f}6$.

The $E_8\times E_8$ case is to some extent similar
but has some peculiarities. Consider the class of models
generated by gauge shifts of the form
\beq
\label{shiftsee}
V\ =\ \frac13( 1,\cdots, 1,0, \cdots, 0)\times \frac13 (1,\cdots, 1,0,\cdots,0)
\eeq
with an even number $m_1$ ($m_2$) of $\frac13$ entries in the first (second) $E_8$
and with $m_1+m_2\leq 8$.  Models with appropriate oscillator
moduli in the twisted sector have
$(m_1,m_2)= (0,0)$, $(2,0)$,$(4,0)$, $(2,2)$, $(2,4)$ and $(4,4)$.  The
corresponding  gauge groups and hypermultiplet spectra are given in 
Table~ \ref{tabla4} (recall that some shifts shown in the table can 
be written in the form (\ref{shiftsee}) through $E_8$ lattice automorphisms).
Again, 
all of these models (except for $(m_1,m_2)=(4,4)$) 
do not fulfill the perturbative modular invariance constraints and 
are also anomalous. However, unlike the $SO(32)$ case, they  
{\it  do not have  
non-Abelian gauge anomalies}.
We can check that they miss an equivalent of $3(8-m_1-m_2)\times 30$ 
hypermultiplets in order to cancel gravitational anomalies. But this is precisely 
the contribution corresponding to the presence in the spectrum of
$3(8-m_1-m_2)$ tensor multiplets and the same number of hypermultiplets.
These missing modes match the non-perturbative spectrum  
corresponding to setting this same number of instantons to zero size
in $E_8\times E_8$. 
This is a nice check of  our  procedure  
since the simple addition of a shift in the vacuum energy 
automatically takes into account
the difference between the $SO(32)$ and
$E_8\times E_8$ heterotic strings, yielding no gauge anomalies in the
second case.
The $Z_3$ models under consideration  are  orbifold analogues of
the $E_8\times E_8$ vacua in the presence of wandering branes considered
in refs.~\cite{dmw,sw6d}. In fact, if we use the hypermultiplets
to Higgs the theory as much us possible we find that the final massless
spectra corresponds to that found in smooth (fully Higgsed)  
$K3$ compactifications with instanton numbers $(k_1,k_2)$, as given in
the final column of Table~\ref{tabla4}, and  $(24-k_1-k_2)$  M-theory five-branes.

\begin{table}[p]
\footnotesize
\renewcommand{\arraystretch}{1.25}
\begin{center}
\begin{tabular}{|c|c|c|c|c|}
\hline
Shift $V$ & & & $E_B$ & \\
\cline{1-1}\cline{4-4}
Group & \raisebox{2.5ex}[0cm][0cm]{Untwisted matter} &
\raisebox{2.5ex}[0cm][0cm]{Twisted matter} & $n_B$ &
\raisebox{2.5ex}[0cm][0cm]{$(k_1,k_2)$} \\
\hline\hline
$(0, \cdots, 0)\times (0, \cdots, 0)$ &
2(1,1) &
18(1,1)$^*$ & $\frac49$ & (0,0) \\
\cline{1-1}\cline{4-4}
$E_8\times E_8$ & & & 24 & \\
\hline\hline
$\frac13(1,1,0, \cdots, 0)\times (0 \cdots, 0)$ &
(56,1)+3(1,1)  &
9(1,1)+ 18(1,1)$^*$  & $\frac39$ & (6,0) \\
\cline{1-1}\cline{4-4}
$E_7\times U(1)\times E_8$ &  &  & 18 & \\
\hline\hline
$\frac13(2, \cdots,0)\times (0, \cdots, 0)$ &
(14,1)+(64,1) &
9(14,1)+18(1,1)$^*$  & $\frac29$ & (12,0) \\
\cline{1-1}\cline{4-4}
$SO(14)\times U(1) \times E_8$ & + 2(1,1)& & 12 & \\
\hline\hline
$\frac13(1,1,0, \cdots, 0)\times \frac13(1,1,0 \cdots, 0)$ &
(56,1) + (1,56)   &
18(1,1)+18(1,1)$^*$  & $\frac29$ &(6,6) \\
\cline{1-1}\cline{4-4}
$E_7\times U(1)\times E_7\times U(1)$ & + 4(1,1) &  & 12 & \\
\hline\hline
$\frac13(1,1,0, \cdots, 0)\times \frac13(2,0 \cdots, 0)$ &
(56,1) + (1,14)   &
9(1,14) + 9(1,1)  & $\frac19$ &(6,12) \\
\cline{1-1}\cline{4-4}
$E_7\times U(1)\times SO(14)\times U(1)$ & + (1,64) + 3(1,1) & + 18(1,1)$^*$  & 6 & \\
\hline\hline
$\frac13(2, \cdots,0)\times \frac13(2,0, \cdots, 0)$ &
(14,1)+(64,1) + &
9(14,1)+9(1,14) & $0$ &  (12,12)  \\
\cline{1-1}\cline{4-4}
$SO(14)\times SO(14) \times U(1)^2$ & (1,14) + (1,64) + 2(1,1)& 
+ 18(1,1)$^*$  &  0  & \\
\hline\hline
$\frac13(2, \cdots,0)\times \frac13(2,0, \cdots, 0)$ &
(14,1)+(64,1) + &
9(1,1) & $\frac39$ & -- \\
\cline{1-1}\cline{4-4}
$SO(14)\times SO(14) \times U(1)^2$ & (1,14) + (1,64) + 2(1,1)& & $Sp(1)^9$ & \\
\hline
\end{tabular}
\end{center}
\caption{Non-Perturbative $Z_3$, $E_8\times E_8$, orbifold models. 
$E_B$ denotes the extra vacuum energy shift. $n_B$ refers to the 
number of M-theory  five-branes needed to cancel anomalies. The last example, 
discussed in Chapter 5, requires
different dynamics involving the presence of extra 
gauge group and charged hypermultiplets. The next to last example is 
modular invariant, but has been included for comparison with the last 
one, since they are connected by a transition.
\label{tabla4} }
\end{table}

A few comments on anomaly cancelation are in order.
In the $E_8\times E_8$ case, although the pure quartic gauge anomaly vanishes, 
there is no full factorization of the total anomaly, as expected.
In the anomaly polynomial there is a factorized piece which is canceled
by the exchange of the usual  perturbative  tensor multiplet  and an extra
piece of the form  $n_B(F^2_1-F^2_2)^2$ which is canceled 
\cite{sw6d, sagn} by the
contribution of the $n_B$ tensor multiplets coming from the five-branes.
Concerning $U(1)$ anomalies, they do not factorize, neither in $SO(32)$ nor in
$E_8\times E_8$ models.  In fact they are spontaneously broken much in the 
same way as in the models with anomalous $U(1)$'s of 
refs.~\cite{gsw, afiu, berkooz}. In contrast,  one can check that 
the $U(1)$'s appearing in modular invariant perturbative 
orbifolds do always have standard factorization.

To end this section we will describe {\it smooth} $K3$ compactifications
with the same kind of spectra as the non-perturbative orbifolds with
embedding (\ref{shifts}). Consider then instantons in the
$U(1)$ background in $SO(32) \supset SU(m) \times SO(32-2m) \times U(1)$.
This $U(1)$ has generator $Q=\frac1{\sqrt{2m}}(1,\cdots,1,0,\cdots,0)$, with
$m$ non-zero entries. The massless spectrum resulting from embedding $k$
instantons in this $U(1)$ follows from the index theorem \cite{gsw, afiu}.
We find
\beqa
& {} & \hspace*{-1.5cm}
(\frac{2k}m - 1) [({\bf \frac{m(m-1)}2}, {\bf 1}, \frac2{\sqrt{2m}}) +
({\bf \ov{\frac{m(m-1)}2}}, {\bf 1}, -\frac2{\sqrt{2m}}) ]  \nonumber \\
& \hspace*{-1.5cm} + & \hspace*{-0.8cm}
(\frac{k}{2m} - 1) [({\bf m}, {\bf 32 -2m}, \frac1{\sqrt{2m}}) +
({\bf \ov{m}}, {\bf 32 - 2m}, -\frac1{\sqrt{2m}}) ] + 
20({\bf 1}, {\bf 1}, 0) 
\label{k3q}
\eeqa
Neglecting $U(1)$ charges, we then see that for $k=3m$ eq.~(\ref{k3q})
reduces to
\beqa
& {} & [({\bf m}, {\bf 32 -2m}) + ({\bf \frac{m(m-1)}2}, {\bf 1}) + 
2({\bf 1}, {\bf 1})] \nonumber \\
& + & 9[({\bf \frac{m(m-1)}2}, {\bf 1}) + 2({\bf 1}, {\bf 1})]
\label{k3orb}
\eeqa
This is precisely the perturbative untwisted plus twisted content of 
the non-modular invariant orbifolds that we have considered. Notice also
that the number of required five-branes is $n_B=3(8-m)$ as we found
before. A similar exercise can be carried out for the two remaining 
embeddings.

\subsubsection{Orbifolds with Wilson Lines}

A larger class of similar models, both for
$E_8\times E_8$ and $SO(32)$,  can be obtained if there are additional
quantized Wilson lines \cite{orbi}.  
The corresponding action on the gauge degrees of freedom 
can be represented by shifts $a_i$, $i=1,2$, where  $3a_i$ belongs to the
$E_8\times E_8$  or $Spin(32)/Z_2$ 
lattice and $i$ labels the two $T^2$ tori in $T^4=T^2\times T^2$.
Now, each of the fixed points has associated to it one of
the 9 possible  shifts $V'=V+n_1a_1+n_2a_2$ with $n_1,n_2=1,0,-1$.
In order to get a model with massless oscillators in all twisted sectors  
all shifts $V'$ have to obey similar constraints as before, although now
the shift $E_B$ in each twisted sector will be different.  Two 
$E_8\times E_8$ examples that require respectively the addition of 8 and 4  
five-branes are described in Tables~\ref{tabla5} and \ref{tabla6}. In the 
first example three of the fixed points have magnetic charge $Q_f=-\frac43$ and 
the other six have $Q_f=-\frac23$. In the second model three of the fixed points have 
$Q_f=-\frac43$ and the other six $Q_f=0$. Thus, in this second example only three of
the fixed points cause the non-perturbative phenomena.
Notice that the total number of  five-branes required in this class of models
is given by
\beq
\label{ello}
n_B\ =\  - \sum _{f}  Q_f  \ =\  6\sum _{f} E_B(f)
\eeq
where the sum goes over all fixed points. In this class of models one can get
theories with {\it any} even number of five-branes from  0 to 24.

\begin{table}[htb]
\renewcommand{\arraystretch}{1.2}
\begin{center}
\footnotesize
\begin{tabular}{|c|c|c|c|}
\hline
\multicolumn{2}{|c|}{$V=\frac13(1,1,0 \cdots,0)\times \frac13(1,1,0,\cdots ,0)$}
 & \multicolumn{2}{|c|}{$a_1=\frac13(0,\cdots,0,1,1)\times (0,\cdots ,0)$} \\[0.2em]
\hline
\multicolumn{4}{|c|}{$SO(12)\times U(1)^2\times E_7\times U(1)$} \\
\hline\hline
Sector & Matter & $Q_f$  & $E_B$   \\
\hline
\hline
 Untwisted &  (32,1)+(1,56)+4(1,1) 
  &   --  &  -- \\
\hline
$V'=V$
 &  6(1,1)+6(1,1)$^*$
  &  -4/3  & 2/9 
\\
\hline
$V'=V+a_1$ 
&  3(12,1)+9(1,1)+6(1,1)$^*$  &   -2/3  &  1/9
\\
\hline
$V'=V-a_1$ 
&  3(12,1)+9(1,1)+6(1,1)$^*$  &   -2/3  &  1/9
\\
\hline
\end{tabular}
\end{center}
\caption{ A $Z_3$ non-perturbative $E_8\times E_8$
orbifold model with one Wilson line.
Anomaly cancelation requires the presence of 8 five-branes.
\label{tabla5} }
\end{table}

\begin{table}[htb]
\renewcommand{\arraystretch}{1.2}
\begin{center}
\footnotesize
\begin{tabular}{|c|c|c|c|}
\hline
\multicolumn{2}{|c|}{$V=\frac13(1,1,0 \cdots,0)\times \frac13(1,1,0,\cdots ,0)$}
 & \multicolumn{2}{|c|}{$a_1=\frac13(0,\cdots,0,2)\times (0,\cdots ,0)$} \\[0.2em]
\hline 
\multicolumn{4}{|c|}{$SO(10)\times SU(2)\times U(1)^2\times E_7\times U(1)$} \\
\hline\hline
Sector & Matter & $Q_f$  & $E_B$   \\
\hline\hline
 Untwisted &  (10,2,1)+(1,1,56)+4(1,1,1) 
  &   --  &  -- \\
\hline
$V'=V$
 &  6(1,1)+6(1,1)$^*$
  &  -4/3  & 2/9 
\\
\hline
$V'=V+a_1$ 
&  3(10,1,1)+3(16,1,1) +6(1,1,1) &   0  &  0 \\
& +3(1,2,1) +6(1,2,1)$^*$ +6(1,1,1)$^*$ & & \\
\hline
$V'=V-a_1$ 
&  3(10,1,1)+3(16,1,1) +6(1,1,1) &   0  &  0 \\
& +3(1,2,1)+6(1,2,1)$^*$ +6(1,1,1)$^*$ & & \\
\hline
\end{tabular}
\end{center}
\caption{ A $Z_3$ non-perturbative 
$E_8\times E_8$ orbifold model with one Wilson line.
Anomaly cancelation requires the presence of 4 five-branes.
\label{tabla6} }
\end{table}

\subsection{Transitions  between Perturbative and Non-Perturbative Orbifold Vacua}

An interesting question is  whether 
there is any shift $V$ in $E_8\times E_8$ or $Spin(32)/Z_2$ 
which admits both spectra with and without 
five-branes.  This is interesting because it can indicate possible
 transitions between perturbative and non-perturbative vacua which
proceed through the emission of five-branes to the $K3$ bulk.
Comparing Tables~ \ref{tabla1}  to \ref{tabla4}   we see that indeed, 
there is a unique case corresponding to the `standard embedding', 
$V=\frac 13(1,1,0,\cdots,0)\times (0, \cdots,0)$
($V=\frac13(1,1,0,\cdots,0)$ for $Spin(32)/Z_2$)  in which 
there are both a model without five-branes and a model with 18 five-branes.
Both models have identical untwisted perturbative spectrum but differ in that
the twisted spectrum  of the perturbative model has  extra hypermultiplets, 
transforming (in the $E_8\times E_8$ case) as 
$({\bf 56},{\bf 1})+7({\bf 1},{\bf 1})$, 
while the non-perturbative one contains just  three  singlets per fixed 
point.  Also, the
fixed points in the non-perturbative model have magnetic charge  $Q_f=-2$. 
This suggests that there can be transitions by which, around
a fixed point in  the perturbative model,  a set of hypermultiplets 
$({\bf 56},{\bf 1})+4({\bf 1},{\bf 1})$ 
is converted into two five-branes which are emitted to the bulk.  The
magnetic charge is conserved in the process since each fixed point has charge 
$Q_f=-2$ and each of the five-branes has charge +1.  
This can happen fixed point by fixed point  so that  there should exist 
similar models to the second model in Table~\ref{tabla4}  with any even number of
five-branes in between 2 and 18.
Thus, in this standard embedding models there is a discrete degree of
freedom which corresponds to having  pairs of zero size instantons.

These are  the unique  heterotic  $Z_3$ models (of the type being considered) 
that admit  both
perturbative and non-perturbative realizations.  Notice that this is related to the
fact that, unlike what happens with the other  $Z_3$  models,  one can
{\it locally} cancel the magnetic charge  in each fixed point by moving
two five-branes into each fixed point. This is not possible in the
other $Z_3$ examples. 
We will see other types of transitions
involving  a Coulomb phase in $SO(32)$ in chapter 5.

We have discussed in this chapter heterotic orbifold models  
rendered consistent by the addition of  five-branes which move in the bulk.
In all these models there are  20  massless  {\it singlets}  corresponding to
the underlying $K3$  moduli  and then the known physics of small instantons in
{\it smooth}  $K3$ compactifications is expected to apply. 
We have not considered shifts of the form  $V=\frac13 (2,1,1,0,\cdots ,0)$.
This case is special since,
unlike the other models, the twisted oscillators 
are not singlets under the non-Abelian gauge group
and their interpretation as $K3$  moduli is not obvious.
In fact, models  obtained by setting $E_B=\frac19$
lack some hypermultiplets to cancel gravitational anomalies.

A natural question is 
what happens with orbifold models in which some moduli are absent. 
This happens  in $Z_3$ 
if the gauge shifts have 
$V^2>\frac89$, as we discussed above.
That also happens in the above orbifold models if, for instance,
we increase the vacuum energy  $E_B\rightarrow E_B + \frac13$.
In this case the singular orbifold cannot be smoothed out  and
some  five-branes can be stuck at the orbifold singularities.
We then need to know the physics of
small instantons at $Z_M$ singularities.
The present understanding of this topic is still incomplete 
(see refs.~\cite{quivers, aspfz2, intri, bi, bi2, mv88}).
Some interesting results are known about  the  existence of new Coulomb 
phases in  $SO(32)$ , $D=6$  theories  when a sufficiently large number of
small instantons sit at an orbifold singularity \cite{intri}.  
Concerning the equivalent  situation in  $E_8\times E_8$, 
results about the non-perturbative enhanced gauge groups
are known in terms of F-theory \cite{mv88}, but we lack as yet sufficient
information about  the hypermultiplets.  
We will show in  Chapter 5
how indeed  the general class of orbifold models  discussed in this chapter
might correspond in some cases to $SO(32)$ heterotic vacua in one of
these Coulomb phases with extra tensor multiplets.

\section{Index Theorems and  Orbifold Singularities}

We would like now to interpret some of the results found in the previous section
in terms of index theorems for instantons on $Z_M$ ALE
spaces.  This will give us a better understanding  of the class of orbifold 
models that we are constructing. In particular, we will see that the 
number of hypermultiplets 
found in the twisted sectors of the orbifolds coincides with the 
dimension of the moduli space of instantons at $Z_M$ singularities,
both for the $Spin(32)/Z_2$ and $E_8\times E_8$ cases.
Another interesting and important question is how to relate the
modular invariance condition (\ref{dos}), or more generally
the shift $V$, with the anomaly conditions
(\ref{anombso}) and (\ref{anombe8}).  

To answer these questions, we consider first the $SO(32)$ heterotic
string on $T^4/Z_M$. We then need to study $SO(32)$ instantons
on orbifold singularities. These instantons are characterized by
an integer $\ell$ and also by a rotation $\Theta$ in $SO(32)$ that takes 
into account the orbifold action. $\Theta$ is related to the shift $V$ by
$\Theta={\rm diag \ }(e^{\pm 2i\pi V_1}, \cdots , e^{\pm 2i\pi V_{16}})$.
The condition $MV \in \Gamma_{16}$ implies $\Theta^M=\pm1$. Indeed,
when $MV$ is a root weight of $\Gamma_{16}$, $\Theta^M=1$ and the instantons have
vector structure, according to the terminology of ref.~\cite{berkooz}.
When $MV$ is instead a spinor weight of  $\Gamma_{16}$, $\Theta^M=-1$ and
the instantons do not have vector structure.

For instantons with vector structure the eigenvalues of $\Theta$ are of the
form $e^{2\pi i \mu/M}$, with $\mu=0,1,\cdots, M-1$. We define
$\wt{w}_\mu$ as the number of such eigenvalues. Notice that 
$\wt{w}_\mu=\wt{w}_{M-\mu}$ and $\sum_\mu \wt{w}_\mu = 32$. 
Then, the instanton number at a $Z_M$ fixed point turns out to be \cite{intri}
\beq
I_f = \ell + \sum_{\mu=0}^{M-1} \ \frac{\mu(M-\mu)}{4M} \wt{w}_\mu
\label{infix}
\eeq
We will also find convenient to define $w_\mu$ as the number of entries 
equal to $\frac{\mu}M$ in $V$. Clearly, $\wt{w}_0 = 2w_0$, $\wt{w}_\mu = w_\mu$
for $\mu < P$ and $\wt{w}_P = 2w_P$ ($\wt{w}_P = w_P$) for $M=2P$ ($M=2P+1$).

The result in eq.~(\ref{infix}) can also be written as
\beq
I_f = \ell + M E_\Theta
\label{inteta}
\eeq
where
\beq
E_\Theta = \sum_{I=1}^{16} \ \frac12 V_I (1 - V_I)
\label{eteta}
\eeq
This $E_\Theta$ is the twisted vacuum energy associated to $\Theta$.
On the other hand, the Bianchi identity  $dH=\tr F^2- \tr R^2$, 
integrated around a fixed point yields
\beq
Q_f \ =\int dH \ =\  I_f  \ -\ C_2(\varepsilon_M)
\label{cargilla}
\eeq
where  $C_2(\varepsilon_M)=(M^2-1)/M$  is the Euler number 
associated to the ALE space $\varepsilon_M$. The latter can be written in 
terms of the vacuum energy $E_{\theta }=(M-1)/M^2$,  corresponding to the 
four compactified bosonic coordinates, as $C_2(\varepsilon_M)=M(M+1)E_{\theta }$.
Then one can finally write
\beq
Q_f\ =\   \ell^{\prime} \ + M(E_{\Theta } \ - \ E_{\theta })
\label{cargolla}
\eeq
with $\ell^{\prime}=\ell-M+1$.  For $Q_f=0$ one recovers
the usual modular invariance constraints of perturbative orbifolds, whereas
for a non-vanishing  $H$-flux at the fixed points one gets  a modified level
matching constraint, as we advanced in the previous chapter.

Given $I_f$, we can  also  compute the total instanton number as
\beq
k = \sum_f \ I_f
\label{totk}
\eeq
It is straightforward to verify that
the modular invariance constraint (\ref{dos}) on $V$  implies 
the condition that the total instanton number adds to 24, for some
integer $\ell$. To this end, recall that $T^4/Z_2$ has 16 $Z_2$ fixed
points; $T^4/Z_3$ has 9 $Z_3$ fixed points; $T^4/Z_4$ has 4 $Z_4$ fixed
points plus 6 $Z_2$ fixed points; and $T^4/Z_6$ has 1 $Z_6$ fixed
point, 4 $Z_3$ fixed points and 5 $Z_2$ fixed points. 

In a general model (modular invariant or not), the total instanton number 
(\ref{totk}) satisfies (\ref{anombso}). For given $\ell$ and $V$
we can then determine the number of allowed five-branes. 
To illustrate the foregoing discussion we will focus on the $Z_3$ orbifold.
In this case, an embedding with vector structure has $m$ eigenvalues
$e^{\pm 2\pi i/3}$ so that
\beq
I_f = \ell + \frac{m}3
\label{infix3}
\eeq
The total instanton number is then
\beq
k=9\ell + 3m
\label{totk3}
\eeq
The condition $k=24$ is satisfied for $m=2,5,8,11,14$ with $\ell=2,1,0,-1,-2$,
respectively. The corresponding shifts are precisely those given in 
Table~\ref{tabla2}. 

We can also consider the models studied in Section 3 having shift (\ref{shifts})
and $n_B=24-3m$. From (\ref{totk3}) and (\ref{anombso}) it follows that these
models have $\ell=0$ and necessarily $m \leq 8$, since otherwise
they would have $k>24$ (for $\ell=0$). Notice that this bound also appeared in 
the orbifold construction, but for different reasons.

It is also possible to give general results for the number of massless 
hypermultiplets at a $Z_M$ fixed point with instanton number $I_f$. This number
is related to the dimension of the moduli space of instantons, 
$\cam_{inst}(M)$, and follows from the index theorem in \cite{aps} for 
manifolds with boundary (the case of $Z_M$ ALE spaces, for which the 
boundary is at infinity is  studied in \cite{roma}). 
For embeddings with vector structure it is found that \cite{intri}
\beq
{\rm dim}\ \cam_{inst}(M) = 30I_f + \frac12[\sum_{\mu,\nu=0}^{M-1}
\wt{w}_\mu \wt{w}_\nu X_{\mu\nu} -  \sum_{\mu =0}^{M-1} \wt{w}_\mu  X_{\mu,M-\mu} ]
\label{dimin}
\eeq
where $X_{\mu\nu}$ is defined by
\beq
X_{\mu\nu} = - \frac1{4M} |\mu - \nu| (M- |\mu - \nu|)
\label{xmunu}
\eeq
At a $Z_M$ ALE space there are also $(M-1)$ blowing-up modes that must be taken
into account.

Let us again illustrate these results for the $Z_3$ orbifold. For $\Theta$
with $m$ eigenvalues $e^{\pm 2\pi i/3}$, the total number of states at a
fixed point, denoted $\cn_f(3)$, is given by
\beq
\cn_f(3) = 30\ell + \frac12 m(m-1) + 2
\label{nfix3}
\eeq
Notice that for the modular invariant embeddings with $m=2,5,8,11,14$
and $\ell=2,1,0,-1,-2$, $\cn_f(3)$ agrees with the number of states per 
fixed point given
in the third column of Table~\ref{tabla2}. Moreover, notice that for 
$\ell=0$, eq.~(\ref{nfix3}) agrees precisely with the number of massless states,
{\it cf.} eq.~(\ref{tspec}), derived from the mass formula with an extra
term $E_B=\frac{(8-m)}{18}$. 

We now give a derivation of the result for ${\rm dim}\ \cam_{inst}$, for
a particular $Z_3$ embedding, starting directly from the index theorem
on a $Z_3$ ALE space $\varepsilon_3$. This exercise will show how to
generalize to the $E_8\times E_8$ case. Consider then,
as in Section~\ref{nmihm},  
instantons in the
$U(1)$ background in $SO(32) \supset SO(32-2m)\times SU(m) \times U(1)$.
As we have seen,
this $U(1)$ has generator $Q=\frac1{\sqrt{2m}}(1,\cdots,1,0,\cdots,0)$, with
$m$ non-zero entries. In the decomposition of the adjoint we find $m(32-2m)$
states of charge $q=\pm \frac1{\sqrt{2m}}$ and $\frac{m(m-1)}2$ states
of charge $q=\pm \frac2{\sqrt{2m}}$. These states give rise to charged
hypermultiplets whose number can be computed from the index theorem. In
general, the number of hypermultiplets of charge $q$, denoted $\cad(q)$,
is given by
\beq
\cad(q) = -\frac1{24} C_2(\varepsilon_3) + q^2I_f + \oh \xi_{\oh}(q)
\label{indge}
\eeq
Here $C_2(\varepsilon_3)=\frac83$ is the Euler number of $\varepsilon_3$
and $I_f$ is given in eq.~(\ref{infix3}). The boundary correction $\xi_{\oh}(q)$
is given by
\beq
\xi_{\oh}(q) = \frac13 \sum_{j=1}^2 \frac{e^{2\pi i \sqrt{2m} \, q j/3}}
{2(1-\cos \frac{2\pi j}3)}
\label{xicor}
\eeq
where the phase in the numerator is such that a state with charge 
$q=\frac1{\sqrt{2m}}$ picks up a phase $e^{2\pi i/3}$ while going once
around the fixed point.
We readily obtain $\xi_{\oh}(\pm \frac1{\sqrt{2m}})=-\frac19$, 
$\xi_{\oh}(\pm \frac2{\sqrt{2m}})=-\frac19$. Therefore,
\beq
\begin{array}{ccccc}
\cad(\frac1{\sqrt{2m}}) & = & \cad(-\frac1{\sqrt{2m}}) & = & \frac{\ell}{2m} 
\nonumber \\
\cad(\frac2{\sqrt{2m}}) & = & \cad(-\frac2{\sqrt{2m}}) & = & 
\frac{2\ell}{m} + \oh \\
\end{array}
\label{dqs}
\eeq
The total number of states is then
\beq
\frac{\ell}{m}[m(32-2m)] + (\frac{4\ell}{m} + 1)[\frac{m(m-1)}2] = 
30\ell + \frac{m(m-1)}2
\label{nfix3again}
\eeq
in agreement with eq.~(\ref{nfix3}), once the two blowing-up modes are included.

The equivalent formulae for the $E_8\times E_8$ case are not
available in the literature so we will repeat the 
 same analysis for $U(1)$ instantons in $E_8$.
Consider for instance the $U(1)$ in $E_8 \supset E_7 \times U(1)$ with
generator $Q_1=\oh(1,1,0,\cdots,0)$. In the adjoint decomposition we find 56 states
of charge $q=\pm \oh$ and one state of charge $q=\pm 1$. 
Since $I_f=\ell_1 + \frac23$ in this case, from eq.~(\ref{indge})
we then find $\cad(\pm \oh) = \frac{\ell_1}4$ and  $\cad(\pm 1) = \ell_1 + \oh$. 
Hence
\beq
{\rm dim}\ \cam_{inst} = \frac{\ell_1}2 \times 56 + 
(\ell_1 + \oh)\times 2 = 30\ell_1 + 1
\label{nfixe8m2}
\eeq
We must also include the two blowing-up modes of the $Z_3$ fixed point.
Taking $\ell_1=2$ we then reproduce the number
of twisted states in the modular invariant standard embedding. On the other
hand, taking $\ell_1=0$ we recover the number of states in the non-perturbative
$Z_3$ model with the same embedding, $E_B=\frac13$, and $n_B=18$, since
$9I_f=6$. Moreover, if we embed the same $U(1)$ in both $E_8$'s, we 
have $I_f=\ell_1 + \ell_2 + \frac43$ and from the index theorem we conclude
${\rm dim}\ \cam_{inst} = 30\ell_1 + 30\ell_2 + 2$, since there are no states
in the $E_8\times E_8$ adjoint with mixed charges. Setting $\ell_1=\ell_2=0$
gives the same number of states found from the mass formula with $E_B=\frac29$.
Also, in this case $9I_f=12$ so that 12 five-branes are needed. 

It is a simple exercise to work out the number of states for other embeddings.
For example, the $U(1)$ with generator $Q_1=\frac1{2\sqrt2}(1,1,1,1,0,\cdots,0)$
breaks $E_8$ to $SO(14)\times U(1)$ and gives 14 states with charge
$q=\pm \frac1{\sqrt2}$ plus 64 states with charge $q=\pm \frac1{2\sqrt2}$.
Using $I_f=\ell_1 + \frac43$ and the index formula (\ref{indge}) we find
\beq
{\rm dim}\ \cam_{inst} = \frac{\ell_1}8 \times 128 + 
\frac{\ell_1+1}2 \times 28 = 30\ell_1 + 14
\label{nfixe8m4}
\eeq
The embedding $Q_1=\frac1{2\sqrt2}(2,0,\cdots,0)$ leads to the same 
results, as it should be due to equivalences in the $E_8$ lattice.
Remembering to include the two blowing-up modes of the $Z_3$ fixed point, we
readily recover the results for the third model in Table~\ref{tabla4} that
has $\ell_1=0$ and $9I_f=12$ so that $n_B=12$. The perturbative model with
the same embedding in both $E_8$'s is also reproduced taking $\ell_1=\ell_2=0$. 

The results reported in this section give the hypermultiplet number in 
the twisted sectors of all models in Tables~\ref{tabla1} to \ref{tabla4}.
Notice, however, that the orbifold analysis gives not only the 
number of  hypermultiplets but their quantum numbers with respect to 
the perturbative gauge group. In addition, the orbifold construction 
gives the spectrum of the complete model including untwisted hypermultiplets,
vector, tensor and gravity multiplets.

\section{Branes at Fixed Points and Tensor Multiplets}
\label{tensors}

Up to this point we have constructed orbifold models that contain enough 
blowing-up 
modes in their perturbative spectrum to smooth out the singular points 
completely. This feature served us as a guide to obtain the relevant
non-perturbative effects rendering the theory consistent. However, as 
mentioned at the end of Section 3.3, one frequently encounters models not 
containing these 
blowing-up modes, and for which the addition of tensors or $Sp(n_B)$ enhanced 
symmetries in the prescribed way does not cancel the anomalies completely.

A natural possibility is that in these models the positions of some or 
all of the five-branes are trapped at the (non-removable) singular points of 
the variety, so that the brane dynamics differs from that found at smooth points, 
and their low energy excitations are in this sense exotic. Our purpose in this 
section is to apply some known results about branes at singularities to the 
understanding of other families of non-perturbative orbifolds.

The behaviour of the $SO(32)$ heterotic five-branes near orbifold points can 
be extracted from the recent studies of type I D-five-branes on ALE spaces 
\cite{quivers, intri, bi,bi2}, and some ideas borrowed from the $Z_2$ case, 
extensively analyzed from the F-theory point of view in \cite{aspfz2}. 
In Table~\ref{tablex} we show the spectrum of some worldvolume theories of
$SO(32)$ five-branes at $Z_M$ singularities which we will 
need in the rest of the article
\footnote{As noticed  in \cite{aspfz2}, the notion of vector 
structure is ill defined along the tensor Coulomb branch. We follow 
ref.~\cite{intri} in relating the existence of a vector structure to the 
gauge shift, as explained in section 2.}.

\begin{table}[phtb]
\renewcommand{\arraystretch}{1.25}
\begin{center}
\footnotesize
\begin{tabular}{|c|c|c|c|}
\hline
$ \! Z_M \! $   &   Gauge Group   &   Hypermultiplets & $\! n_T \! $ \\
\hline
\hline
\multicolumn{4}{|c|}{Embeddings with vector structure 
} \\ 
\hline -- &  $Sp(\ell)$  &   ${\frac {32}2}(2\ell)+(\ell(2\ell-1))$ & 0 \\
\hline
$Z_2$  &   $Sp(\ell)\times Sp(\ell+{\frac {w_1}2}-4) $  
 &  $w_0(2\ell,1)+w_1(1,2\ell+w_1-8)+(2\ell,2\ell+w_1-8)$ & 1 \\
\hline
$Z_2$ & $Sp(\ell)\times SO(2\ell+8) \quad [w_1=0]$ & $(2\ell, 2\ell+ 8)$ & 1 \\
\hline
$Z_3$  & $Sp(\ell)\times U(2\ell+w_1-8)$    
&    $w_0(2\ell,1)+w_1(1,2\ell+w_1-8)$ & 1 \\
 &   &  $ + (2\ell, 2\ell+w_1-8)+(1,(\ell+{\frac{w_1}2}-4)(2\ell+w_1-9))$ & \\
\hline
$Z_4$  &   $Sp(\ell)\times U(2\ell+w_1+w_2-8) $  &
  $w_0(2\ell,1,1)+ w_1(1, 2\ell+w_1+w_2-8,1)$ & 2 \\
              &    $\times  Sp(\ell+\frac{w_1}2+w_2-8)$  &
$+w_2(1,1,2\ell+w_1+2w_2-16)+(2\ell,2\ell+w_1+w_2-8, 1) $  & \\
             &       & $+(1, 2\ell+w_1+w_2-8,  2\ell+w_1+2w_2-16)$ & \\
\hline\hline
\multicolumn{4}{|c|}{Embeddings without vector structure} \\
\hline
$Z_2$  &   $U(2\ell)$  &  ${\frac {32}2}(2\ell)+2(\ell(2\ell-1))$ & 0 \\
\hline
$Z_4$  &   $U(2\ell)\times U(2\ell+u_2-8)$    &
   $u_1(2\ell,1)+u_2(1,2\ell+u_2-8)+ (2\ell, 2\ell+u_2-8)$ & 1 \\
    &   &   $ + (1,(\ell+{\frac{u_2}2}-4)(2\ell+u_2-9))+(\ell(2\ell-1),1)$ & \\
\hline
$Z_6$  &   $U(2\ell)\times U(2\ell+u_2+u_3-8)$    &
$u_1(2\ell,1,1)+u_2(1,2\ell+u_2+u_3-8,1)  + (\ell(2\ell-1),1,1)  $  & 2 \\
     & $\times U(2\ell+u_2+2u_3-16)$  &  
           $+u_3(1,1,2\ell+u_2+2u_3-16)+(2\ell,2\ell+u_2+u_3-8,1)$ & \\
   &   & $+ (1,2\ell+u_2+u_3-8,2\ell+u_2+2u_3-16)  $ & \\
      &     & $+(1,1,(\ell+{\frac{u_2}2}+u_3-8)(2\ell+u_2+2u_3-17))$ & \\
\hline
\end{tabular}
\end{center}
\caption{ Some world-volume theories of $SO(32)$ five-branes at $Z_M$ singularities.  
Here $w_\mu$ is the number of entries equal to $\frac{\mu}M$ in $V$ with
vector structure. Similarly, $u_\mu$ is the number of entries equal to 
$\frac{2\mu -1}{2M}$ in $V$ without vector structure.
\label{tablex}}
\end{table}

Specifically, in \cite{intri} it was argued (based on the analysis of 
\cite{quivers} and anomaly considerations, further confirmed by a 
detailed determination of world-sheet consistency conditions \cite{bi}) 
that when a large enough number $\ell$ of five-branes sit on a $Z_3$ 
singular 
point (with possible vector structure), a $Sp(\ell) \times U(2\ell+m-8)$ 
gauge symmetry develops with hypermultiplets transforming as 
\beqa
& {} & (16-m) ({\bf 2\ell},{\bf 1}) + ({\bf 2\ell},{\bf 2\ell+m-8}) 
+ m({\bf 1},{\bf 2\ell+m-8}) 
\nonumber \\
& {} & \hspace*{2cm} + 
({\bf 1},{\bf (\ell+\frac{m}2-4)(2\ell+m-9)})
\label{npin}
\eeqa 
The multiplicity can be understood as gauge quantum numbers under the 
perturbative symmetry group.
Furthermore, there also appears an extra
tensor degree of freedom. It is also stressed that on the Coulomb 
branch parametrized by the scalar in this tensor multiplet, one of the two 
perturbative blowing-up modes is absent, since the 
singular point cannot be completely smoothed out while preserving the 
tensor multiplet in the spectrum. Thus the possibility of understanding 
perturbative spectra with missing blowing-up modes opens up.

Note that this world-volume theory 
makes sense as long as $\ell \geq \frac{8-m}2$, thus a minimum value on 
$\ell$ is required for having the singular point on the Coulomb phase. 
The $m=8$ case is special, since the transition to the Coulomb 
branch is possible even for $\ell=0$. The described spectrum reduces to just 
one 
tensor multiplet, without any gauge enhancement. Remarkably enough, this 
is precisely the non-perturbative contribution which is required to 
complete the $Z_3$ orbifold with shift 
$V=\frac13(1,1,1,1,1,1,1,1,0,\cdots,0)$ and $E_B=\frac13$. Its spectrum is 
shown  at the bottom of 
 Table~\ref{tabla3}, and we see that each twisted sector contributes 
with a singlet. Once the extra tensors are added, the model is 
anomaly-free. Observe that it also reproduces the $Z_3^A$ orientifold of 
ref.~\cite{gj, dabol1}. It was noted in \cite{intri} that
the origin of the tensors in this  orientifold could be understood in terms of 
such a Coulomb phase.
Our orbifold construction, on the other hand, yields 
a global matching with the orientifold spectrum, including the untwisted 
and twisted sectors, thus providing the complete heterotic dual.  

Notice that there exists a modular invariant perturbative orbifold with 
this same shift $V$, as shown in Table~\ref{tabla3}. Each fixed point 
contributes with one ${\bf 28}$ of $U(8)$ and two blowing-up modes, in 
agreement with the number of states obtained from the index theorem 
(\ref{nfix3}). This 
model corresponds to having the singular points in the Higgs phase. The 
transition 
to the Coulomb branch is dominated by tensionless strings in analogy 
with the familiar zero size $E_8$ instanton \cite{gh,sw6d}, and in the 
process the spectrum at a fixed point changes as
\beq
{\bf {28}} + {\bf {1}} \longrightarrow {\rm tensor,}
\label{ttran}
\eeq
 which is consistent with anomaly cancelation conditions, because the 
 28 does not contribute to  the pure quartic gauge anomalies. The transition 
can occur locally at each fixed point, so that there is a whole family of models 
with the number of tensors $n_T$ varying from zero to nine.

One can easily construct a further class of $Z_3$ models with $m\leq 8$ 
and singular points in the Coulomb phase. Consider a $Z_3$ twisted sector 
feeling a shift $V=\frac13(1,\cdots,1,0,\cdots,0)$ with an even number $m$ of
$\frac13$ entries. Including an extra energy shift $E_B=\frac{14-m}{18}$ 
leads to the required perturbative spectrum, namely one singlet per fixed point,
playing the role of the surviving blowing-up mode. However, note that in 
order to reach the Coulomb branch at least 
$\frac{(8-m)}2$ five-branes should be located at each fixed point, and 
this cannot be done for all of them, due to the bound of 24
for the total instanton number. Hence, one is 
forced to consider models that contain two kinds of fixed points. All 
share the same gauge shift, but a number $9-r$ of them have 
$E_B=\frac{8-m}{18}$ (Higgs phase points) and the remaining $r$ have 
$E_B=\frac{14-m}{18}$ (Coulomb branch points). A total of $3m$ large 
instantons is located at the fixed points, $\ell_i$ small instantons are sitting 
at the i-th singularity, and $(24-3m-\sum_{i=1}^r \ell_i)$ five-branes wander 
around the $K3$ bulk. The final spectrum is easily determined by 
adjoining to the orbifold perturbative states (with the corresponding $E_B$ 
at each twisted sector) the massless modes associated to the 
adequate number of trapped and wandering five-branes. The result when the 
$\ell_i$ are set to their critical value $\ell_i=\ell_c=\frac{8-m}2$ 
(at which the non-perturbative unitary group is absent) and 
the $n_B=(6-r)\ell_c$  wandering branes are coincident, is as follows
\beqa
\begin{array}{cl}
U(m) \times SO(32-2m)  \times \prod_{i=1}^r Sp(\ell_c) 
\times Sp(n_B) & \\[0.5em]
\left. \begin{array} {c}
({\bf m},{\bf 32-2m};{\bf 1},\cdots,{\bf 1};{\bf 1})  + 
({\bf \frac{m(m-1)}2},{\bf 1};{\bf 1},\cdots,{\bf 1};{\bf 1}) \\
+2({\bf 1},{\bf 1};{\bf 1},\cdots,{\bf 1};{\bf 1}) 
\end{array} \right\} & {\rm Untw.} \\[0.5em]
\left.\begin{array}{c}
+(9-r) \left[ ({\bf \frac{m(m-1)}2},{\bf 1};{\bf 1},\cdots,{\bf 1};{\bf 1}) 
+ 2({\bf 1},{\bf 1};{\bf 1},\cdots,{\bf 1};{\bf 1}) \right]  \end{array} 
\right\} & {\rm Higgs} \\[0.5em]
\left. \begin{array}{c}
+r({\bf 1},{\bf 1};{\bf 1},\cdots,{\bf 1};{\bf 1}) + \frac 12 
({\bf 1},{\bf 32-2m};\underline{{\bf 1},2\ell_c,\cdots,{\bf 1}};{\bf 1}) 
+ r \ {\rm tensors} \end{array} \right\} & {\rm Coulomb} \\[0.5em]
\left. \begin{array}{c}
+({\bf m},{\bf 1};{\bf 1},\cdots,{\bf 1};{\bf 2n_B}) + \frac 12 ({\bf 
1}, {\bf 32-2m};{\bf 1},\cdots,{\bf 1};{\bf 2n_B}) \\
+({\bf 1},{\bf 1};{\bf 1},\cdots,{\bf 1};{\bf n_B(2n_B-1)-1}) +({\bf 
1},{\bf 1};{\bf 1},\cdots,{\bf 1};{\bf 1}) \end{array} \right\} & 
{\rm Witten} \end{array}
\label{aa1}
\eeqa
where underlining means permutation.
One can check that all gauge and gravitational anomalies cancel.  
Thus, these two possible $E_B$ in this class 
of models reproduce the two phases of the dynamics of branes at the 
singular point.
In analogy with the $m=8$ case, one can easily follow the change of states 
involved in the transition to the Coulomb branch ($r \rightarrow r+1$).

With the knowledge acquired we can now interpret yet
another class of models, whose fixed points have shifts 
$V=\frac13(1,\cdots,1,0,\cdots,0)$, with a number of nonzero entries $8< m 
\leq 14$, and $E_B=\frac{14-m}{18}$. The formula for the vacuum energy 
shift as function of the monodromy twist, and the resulting perturbative 
spectrum, just a singlet per fixed point, suggest that this kind of fixed 
points are frozen 
at the Coulomb branch and should be completed using the non-perturbative 
content described at the beginning of this section. As happened before, we 
must face a global 
subtlety due to the constraint on the total instanton number on $K3$. 
Following eq.~(\ref{infix3}), we see that, even if we set $\ell=0$ for all 
the nine fixed points, more than 24 instantons are required. The solution 
consists again in not treating all singular points symmetrically, 
though in this case they will also differ in the shifts they feel, by 
means of the introduction of quantized Wilson line backgrounds.

The simplest such example has 
\beqa
V & = & (0,\cdots,0) \nonumber \\
A & = & \frac13(1,\cdots,1,0,\cdots,0)
\label{ava}
\eeqa
with the Wilson line  
$A$ having $m$ nonzero entries. We have six points feeling the 
shift  $A$ under study and three with trivial monodromy. Let us briefly 
discuss its spectrum. The untwisted sector contains the gauge 
group $U(m)\times SO(32-2m)$ and two singlets. In the twisted sector, 
the three points with shift $V=0$ contribute with two blowing-up singlets 
each, as expected from index theory, while the remaining six points give one 
perturbative singlet each. The 
total instanton number used up to now is $2m$, so the non-perturbative 
spectrum to be added corresponds to the massless modes of $24-2m$ 
wandering five-branes and six fixed points on the Coulomb phase. The final 
spectrum (the wandering branes are taken coincident, for notational 
convenience) is
\beq
\begin{array}{cl}
U(m)\times SO(32-2m) \times  Sp(24-2m)\times U(m-8)^6 & \\[0.5em]
\left. \begin{array}{c}
14({\bf 1},{\bf 1};{\bf 1}; {\bf 1},\cdots,{\bf 1}) \end{array} \right\} 
& {\rm Pert.} \\[0.5em]
\left. \begin{array}{c}
+ ({\bf m},{\bf 1};{\bf 48-4m};{\bf 1},\cdots,{\bf 1})  +
\frac 12({\bf 1},{\bf 32-2m};{\bf 48-4m};{\bf 1},\cdots,{\bf 1}) + \\
({\bf 1},{\bf 1};{\bf \frac{(48-4m)(47-4m)}2 - 1};{\bf 1},\cdots,{\bf 1}) 
+({\bf 1},{\bf 1};{\bf 1};{\bf 1},\cdots,{\bf 1}) 
\end{array} \right\} & {\rm Witten} \\[0.7em]
\left. \begin{array}{c}
+ ({\bf m},{\bf 1};{\bf 1};\underline{{\bf m-8},\cdots,{\bf 1}})  +
({\bf 1},{\bf 1};{\bf 1};
\underline{{\bf \frac{(m-8)(m-9)}2},\cdots,{\bf 1}}) + 6 \ {\rm tensors} 
\end{array} \right\} & {\rm Coulomb}
\end{array}
\eeq
All gauge and gravitational anomalies cancel.

Another, more complicated, family of models can be constructed with the 
choice
\beqa
V & = & \frac 13 (\underbrace{1,\cdots,1}_{16-m},0,\cdots,0) \nonumber\\
A & = & \frac 13 (\underbrace{-\oh,\cdots, -\oh}_{16-m}, \oh,\cdots,\oh)
\label{ava2}
\eeqa
with $8\leq m \leq 14$. It contains three sets of twisted subsectors 
(with three fixed points each) feeling the gauge shifts
\beq
\begin{array}{ccccccc}
V & = & \frac13 (\underbrace{1,\cdots,1}_{16-m},0,\cdots,0)  & , &   
E_B & = & \frac{m-8}{18}  \nonumber \\
V+A & = & \frac13 (\oh,\cdots \cdots,\oh) & , &
E_B & = & \frac29  \nonumber \\[0.2ex]
V+2A & = & \frac13 (0,\cdots,0,\underbrace{1,\cdots,1}_{m})  & , & 
E_B & = & \frac{14-m}{18} 
\end{array}
\label{ava3}
\eeq
Thus, three fixed points do not have vector structure (their 
perturbative spectrum is shown in Table~\ref{tabla3}, and has been 
described in Section~3), three have a shift 
with $16-m\leq 8$ nonzero entries (and are in the Higgs branch) and
three have a shift with $m\geq 8$ nonzero entries (so are frozen at the 
Coulomb phase). In that situation the model must be completed 
non-perturbatively with four wandering branes and some Coulomb branch
content, the final result being completely 
free of gauge and gravitational anomalies. The wandering branes can be 
used to put some of the Higgs phase points in the Coulomb phase (a 
process nicely accounted for through a change in the corresponding $E_B$, 
as described above), always leading to consistent results.

We stress that the models we have described are the first global 
constructions in which one can follow the transitions to the Coulomb branch 
due to the piling up of small instantons at singular points. It is 
remarkable that the simple recipe of introducing $E_B$ allows us to 
compute easily not only the number of states being swallowed in that 
transition, but also  their quantum numbers under the perturbative and 
non-perturbative gauge symmetries.

Finally, let us point out that these results concern exclusively the 
$SO(32)$ heterotic orbifolds, since only for them the dynamics of 
five-branes at singular points of $K3$ has been determined
with sufficient  detail. Unfortunately 
such a detailed knowledge is still lacking for the $E_8\times E_8$ heterotic 
non-perturbative effects. However, some information can be extracted from 
the F-theory analysis of \cite{mv88} for the case of unbroken $E_8\times 
E_8$. It can be shown that when four small $E_8$ instantons coalesce on 
top of a $Z_3$ singular point, a non-perturbative $SU(2)$ appears, as well 
as four tensor multiplets, and hypermultiplets transforming as 4({\bf 2})'s. 
As in 
the $SO(32)$ case, one of the two blowing-up modes of the singular points 
disappears. 

It is easy to construct a non-perturbative orbifold in which this proposal 
is realized. Consider for example the $Z_3$ orbifold with $V\!=\!(0,\cdots,0)$ 
and $E_B=\frac49$\, in eight fixed points and $E_B=\frac79$\, in the remaining. The 
perturbative gauge group is $E_8\times E_8$ and there are two untwisted 
moduli. The fixed points with $E_B=\frac49$\, generate two blowing-up modes each, 
while the fixed point with $E_B=\frac79$\, gives just one singlet. The missing 
modes signal the existence of instantons frozen at the singular point. 
Consequently we must add the spectrum just mentioned, associated to four 
$E_8$ instantons at an $A_2$ singularity. The model is finally rendered 
consistent by adding twenty wandering five-branes. We stress that the 
appearance of $E_8$ instantons stuck at singular points is due to the 
shift $E_B\to E_B +\frac13$ in close analogy with the $SO(32)$ case 
(actually, this is consistent  with the equivalence of both theories upon 
compactification to $D=5$ on a further $S^1$ \cite{mv88}).
We also notice that the F-theory version of 
this model is provided by the mirror of the last Calabi-Yau on page 32 of 
ref.~\cite{canpera}, and its interpretation is consistent with the 
conjecture in \cite{mirror}.

It would be interesting to extend this constructions to the case of fixed 
points with nontrivial gauge twists, in order to check that the 
orbifold perturbative 
spectrum is completed by the appropriate non-perturbative contribution. As 
already mentioned, these spectra have not been determined in the literature.
But reversing the viewpoint we can try to extract 
this information precisely by imposing the consistency of our 
non-perturbative orbifolds with missing blowing-up modes, which do not obey 
the usual `tensor + hypermultiplet' rule. 

Consider for example the $E_8\times E_8$ heterotic $Z_3$ orbifold 
defined by the shift $V=\frac 13(1,1,1,1,0,0,0,0)\times 
(1,1,1,1,0,0,0,0)$ and $E_B=\frac13$, whose perturbative spectrum is 
described in Table~\ref{tabla4}. The gauge group is 
$SO(14)^2\times U(1)^2$, the untwisted hypermultiplet content is 
\beq
({\bf 
14},{\bf 1})+ ({\bf 64},{\bf 1})+({\bf 1},{\bf 14})+({\bf 1},{\bf 
64})+2({\bf 1},{\bf 1})
\label{aso14}
\eeq
Also, taking into account the shift $E_B$, each twisted sector contributes 
with just one singlet hypermultiplet. Contrary to the usual  $E_8\times E_8$ 
case, besides gravitational anomalies,
this spectrum presents severe $SO(14)$ gauge anomalies. There are 
nine missing ${\bf 14}$'s of each $SO(14)$ that should arise 
non-perturbatively. A curious solution to the gravitational and gauge 
anomaly cancelation 
conditions is provided by adding a non-perturbative $Sp(1)^9$ gauge group 
and hypermultiplets transforming as $\oh({\bf 14},{\bf 1};\underline{{\bf 
2},\cdots,{\bf 1}}) + \oh({\bf 1},{\bf 14};\underline{{\bf 
2},\cdots,{\bf 1}}) + 2({\bf 1},{\bf 1};\underline{{\bf 2},\cdots,{\bf 1}})$.

The situation is reminiscent of what happens in the $SO(32)$ case. This 
is perfectly sensible, since this orbifold model is associated to a 
heterotic compactification on a smooth $K3$ with $(12,12)$ instantons 
embedded in $E_8\times E_8$, thus $n=0$ in the notation of 
Section~2. This model is known to develop non-perturbative 
gauge symmetries at special loci in its moduli space, as required by 
heterotic/heterotic duality \cite{dmw}, and moreover these effects have 
been interpreted as the shrinking of instantons in the $SO(32)$ T-dual 
version \cite{berkooz}. Thus our choice of additional contributions 
should correspond to this kind of effects.

Note that the analogy with the Witten content is not complete, since the 
proposed spectrum does not contain the two-index antisymmetric 
representation of the $Sp(1)$'s. These singlets would parametrize the 
positions of the (T-dual) five-branes on $K3$, so its absence shows that the 
non-perturbative dynamics is frozen at the fixed points in the orbifold. 
In particular, one cannot make the dual five-branes coalesce, since no gauge 
enhancement is possible without those singlets.
 
Moreover, this model has a modular invariant relative 
with the same untwisted sector, but twisted matter containing the 
required {\bf 14}'s (see Table~\ref{tabla1}), and both are 
connected by Higgsing of the non-perturbative symmetry. Note that in this 
process the ${\bf 64}$'s are not touched (since they live in the 
untwisted sector) and so the dynamics involved does not correspond to the 
shrinking of an $E_8$ instanton \footnote{A $U(1)$ bundle construction 
inmediatly shows that an $E_8$ instanton yields an anomaly-free 
combination of {\bf 14}'s and {\bf 64}'s.}.

In this sense, one encounters again that the procedure of shifting the
energy by $E_B$ 
naturally determines the 
non-perturbative effects relevant to each model.
 This last model presents evidence in 
favour of new dynamics in $E_8\times E_8$ heterotic compactification on 
singular varieties. As was mentioned above, some results on this topic have been
obtained in \cite{mv88} for the case of unbroken $E_8\times E_8$, but a similar analysis 
as symmetry breaking takes place would be required to check the proposed 
spectrum.

\section{Even $M$ $Z_M$  Models and the Heterotic Duals  of Type IIB
Orientifolds}

It is possible to generalize the non-perturbative $Z_3$ orbifold construction
in order to treat $Z_2$, $Z_4$ and $Z_6$ orbifolds.  
However, there are some peculiarities.
To start with, most of the $Z_2$ gauge shifts lead to orbifolds which 
have no  singlets adequate to play the role of $K3$ moduli and
the models obtained often do not correspond to 
smooth compactifications with some zero size instantons  (as happened in the
$Z_3$ models described in Chapter 3). Also, since $Z_4$ and $Z_6$ 
orbifolds have $Z_2$ subsectors, they share the same property.  
Take as an example the $Z_2$ orbifold  in $Spin(32)/Z_2$ 
 with standard embedding and $E_B=\frac12$.
There are no oscillator modes left that could be identified
with $K3$ twisted moduli and the model has hypermultiplets 
in four ${\bf 28}$'s  of $SO(28)$. To cancel gauge anomalies, sixteen ${\bf 
28}$'s
are needed that could come from  sixteen small instantons.  However, since
there are no $K3$ moduli available, there  could  be one small instanton 
(with vector structure)  stuck at each $Z_2$ fixed point. 
The non-perturbative spectrum corresponding to this case has not been 
determined in the literature.  Notice that  in this case the results in Table 
\ref{tablex} are of no use since with $w_1=2$  we need  a minimum of
$l=3$ at  each fixed point  and that would lead to an inconsistent 
vacum with total instanton number bigger than $24$.

There are however some cases where this difficulty is absent, and we 
show below how the  heterotic dual of the 
Bianchi-Sagnotti-Gimon-Polchinski (BSGP)
$Z_2$ orientifold with eight dynamical five-branes at the same
fixed point  can be understood
as a  particular  heterotic $Z_2$ orbifold with small instantons.
Also,  we showed in the previous section how certain orbifold 
heterotic models   without  $K3$ moduli can be understood as
models in a Coulomb phase with extra tensor multiplets.
We will show that this is also the case in some $Z_4$ and $Z_6$
heterotic orbifolds. In the last subsection we construct a $Z_2$ 
orbifold with the same spectrum as orientifold  
models constructed by  Dabholkar and Park \cite{dabol2}
and Gopakhumar and Mukhi \cite{gm}.

\subsection{Heterotic Duals of Type IIB Orientifolds}
\label{dualgj}

To start with, consider the models presented in Table~\ref{tabla7}. 
These are the four $Z_M^A$ (in the notation of ref.  \cite{gj} )
Type IIB orientifolds 
\cite{gj, dabol1}. We will construct the heterotic
duals of all these 
($M$ even) models as  orbifolds,  with 
gauge embedding not verifying the modular invariance constraints,
in the presence of small instantons.

\begin{table}[htb]
\renewcommand{\arraystretch}{1.2}
\begin{center}
\footnotesize
\begin{tabular}{|c|c|c|c|c|c|c|}
\hline
$\! \! Z_M^A \! \! $  &  $\! \! n_T \! \!$
&   (99)-Gauge Group
   &   (99)-Hyper.
 &  $Spin(32)/Z_2$ Shift  & $\! \! E_B(j) \! \!$   &   $\! \! n_H(j) \! \!$   \\
\hline
\hline
 $Z_2^A$
 &  0
  &   $U(16)$  &   $2(120)$  &  $V=\frac14(1,\cdots , 1)$
&  $\frac 14 $    &   16
    \\
\hline
$Z_3^A$
 &  9
  &  $SO(16)\times U(8)$
&   $(1,28)+(16,8)$  &  $V=\frac13(1,1,1,1,1,1,1,1,0,\cdots ,0)$
& $\frac 13 $  &  9
\\
\hline
$Z_4^A$  &  4
&  $U(8)\times U(8)$
 &   $(28,1)+(1,28)$  &
$V=\frac18(1,1,1,1,1,1,1,1,3,\cdots ,3)$ 
&   $\frac 3{16} $    &   4  \\
& & & +(8,8) &  &  $\frac 14 $  &   10  \\
\hline
$Z_6^A$   &  $ 6$
&  $U(4)\times U(4)\times U(8)$
&    $(6,1,1)+(1,6,1)$
 &  $V=\frac1{12}(1,1,1,1,5,5,5,5,3,\cdots ,3)$
&     $\frac14 $   &  2 \\
 & & & +(4,1,8)+(1,4,8) &  & $\frac 13 $  &  5   \\
& & & &  &  $\frac 14 $  &  5 \\
\hline
\end{tabular}
\end{center}
\caption{ Type IIB $Z_M^A$ orientifolds
and their heterotic duals.  
\label{tabla7}}
\end{table}

The second column gives the number of tensor multiplets, whereas the third and
fourth  show the multiplet content in the (99) sectors in the orientifolds.  
The (99) vector multiplets are precisely those expected to be  reproduced from
a perturbative heterotic model.  It is easy to find shifts $V$ with
$NV$ belonging to the $Spin(32)/Z_2$  lattice such that
the untwisted sector of  an heterotic orbifold contains exactly 
these (99)  vector and hypermultiplets. Such shifts for the 
different $Z_M^A$ are shown in the  fifth column of Table~\ref{tabla7}.  
The $Z_3^A$  model was discussed in Chapter~\ref{tensors}, it can be understood
in terms of a transition to a Coulomb phase involving nine tensor multiplets.

The $Z_2^A$ model is the BSGP orientifold \cite{bso, gp}, shown 
to be related to  certain $Z_2$ orbifolds 
of $Spin(32)/Z_2$  and $E_8\times E_8$ \cite {berkooz}. The closed string 
spectrum produces twenty $K3$ moduli. The complete open string spectrum has  a 
$U(16)_{9}\times U(16)_{5}$ gauge group with hypermultiplets in
$2({\bf {120}},{\bf {1}})+ 2({\bf {1}},{\bf {120}})+({\bf {16}},{\bf
{16}})$.  This is the case if all eight dynamical  D-five-branes coincide at
the same  fixed point.  If  half a  five-brane is located at each of the 16 fixed
points, the gauge group is $U(16)_{9}\times U(1)_{5}^{16}$ with
hypermultiplets transforming as 
$2({\bf {120}})+16({\bf {16}})+20({\bf {1}})$. In fact the
$U(1)^{16}_{5}$ is broken and swallows sixteen of the singlet hypermultiplets in a
variation of the Green-Schwarz mechanism \cite{berkooz}.  This is the  particular 
BSGP model that admits a $SO(32)$ heterotic  dual which is a free field
theory \cite{berkooz}.  
Indeed, a standard perturbative (modular invariant)  $Z_2$ orbifold
of $Spin(32)/Z_2$  with shift $V=\frac14(1, \cdots ,1,-3)$ has exactly the
same spectrum (see Table~\ref{tabla2}).  While the $16({\bf {16}})$ hypermultiplets 
of this model originate in a standard (perturbative)
twisted sector, in the orientifold they originate in the (59) open string sector.  
It is also interesting to consider the instanton number and the number
of states at the $Z_2$ fixed points. Since the shift is modular invariant,
we expect $I_f=\frac32$ so that $16I_f=24$. The dimension of $\cam_{inst}$
can be worked out taking into account that $V$ breaks $SO(32)$ to $U(16)$ 
\cite{berkooz}. It is found that ${\rm dim}\ \cam_{inst}=30(I_f -1)=15$.
Including the blowing-up mode gives 16 states per fixed point as expected.

We would like now to obtain the heterotic dual of the particular
BSGP model with gauge group $U(16)_9\times U(16)_5$. Consider 
a $Z_2$ orbifold with $Spin(32)/Z_2$ embedding $V=\frac14(1, \cdots ,1)$. 
This shift, {\it unlike}  that discussed in the previous paragraph,
is not modular invariant by itself since  $2(V^2-\oh)$ is not even.
The  perturbative gauge group is  $U(16)$ with  untwisted
hypermultiplets  $2({\bf {120}})+4({\bf {1}})$. Level matching can
be achieved by adding an extra vacuum energy $E_B=\frac14$ that then leads
to $16({\bf {1}})$ extra hypermultiplets.  

Anomaly cancelation requires the presence of $16({\bf 16})$
hypermultiplets of $U(16)$.
In fact notice that this specific $V$ corresponds to a $Z_2$ without 
vector structure. As discussed in \cite{aspfz2}, in such a situation,  
whenever 8 small instantons coalesce at an orbifold singularity, a $U(16)$ 
non-perturbative gauge group is generated with non-perturbative 
matter content $16({\bf {16}})+ 2({\bf {120}})$. Therefore,
considering both perturbative and non-perturbative contributions,the full BSGP
model is reproduced. Notice that all twenty $K3$ moduli are present here. Moving
the small instantons away from the singularity corresponds to Higgsing down
to $Sp(8)$ by giving a vev to a ${\bf 120}$, leaving a standard small
instanton content. It is consistent to require eight five-branes since in this
case $I_f=1$ as shown in \cite{berkooz}. Also, notice that 
${\rm dim}\ \cam_{inst}=30(I_f -1)=0$, meaning that there is only one
(blowing-up) state per fixed point.

The $Z_4^A$  orientifold  \cite{gj, dabol1}  has a similar structure. 
Consider the non-modular invariant shift
\beq 
V=\frac18(1,1,1,1,1,1,1,1,3,\cdots ,3)
\label{vz4gj}
\eeq
It is easy to check that 
the untwisted sector of the orbifold corresponding to 
this embedding reproduces the (99)-sector
of the  orientifold model. 
By including an energy shift $E_B=\frac3{16}$ it achieves level
matching in the twisted $\theta $ sector and produces four singlets. Another
10 singlets are obtained from the  $\theta ^2$ sector 
with $E_B=\frac 14$ (notice that $2V$
corresponds to the BSGP shift discussed above). Again, this model is a $Z_4$
without vector structure and after adding $E_B(j)$
only one singlet per twisted sector survives. The absence of  some  blowing-up 
modes suggests, in analogy with our experience with $Z_3$, that there is a 
transition to a Coulomb phase in which tensor multiplets appear.
In this case the non-perturbative content can be
read from the results of refs.~\cite{intri, bi}. In order to use
the analysis of instantons at $Z_M$ ALE spaces we must 
take into account that the $Z_4$ orbifold has four $Z_4$ fixed
points and  six  $Z_2$ fixed points. Hence, there are four $Z_4$ ALE spaces
$\varepsilon_4$ and six $Z_2$ ALE spaces $\varepsilon_2$.

The embedding in (\ref{vz4gj}) corresponds to instantons  without vector structure.
{} From refs.~\cite{intri, bi} one learns that in the case of
small instantons on a $Z_4$ singularity there is an enhanced gauge group
\beq
U(2\ell_4)\times U(2\ell_4+u_3-8)
\label{intriblum}
\end{equation}
where $\ell_4$ is an integer large enough to guarantee $(2\ell_4+u_3-8) \geq 0$, 
and $u_3$ is the number of $\frac38$ entries in $V$. Likewise, $u_1=(16-u_3)$ is 
the number of $\frac18$ entries in $V$. In the case at hand $u_3=8$.
There is also one tensor multiplet and hypermultiplets transforming as
shown in Table~\ref{tablex}.
The $Z_4^A$  orientifold 
model is reproduced by setting $\ell_4=4$ at one of the
$Z_4$ fixed points and $\ell_4=0$ at the other three fixed points.
{}From the  four small instantons at the 
first fixed point one gets an $U(8)\times U(8)$ non-perturbative gauge
group with $8({\bf {8}},{\bf {1}})+8({\bf {1}},{\bf {8}}) +({\bf {8}},{\bf
{8}}) +( {\bf {28}},{\bf {1}}) +({\bf {1}},{\bf {28}})$ hypermultiplets and
{\it one tensor multiplet}.  From the other three we get one tensor multiplet 
{} from each and  no enhanced gauge symmetry.  
One can check that the magnetic charge at the six $\epsilon _2$ is
$-\frac12$ whereas that at the four $\epsilon _4$ is $-\frac14$, so that the total
charge coming from the singularities cancels that coming from the 
four small instantons. The given multiplicity of the hypermultiplets 
can now be understood as the gauge quantum 
numbers under the perturbative groups.
Putting together the perturbative spectrum from the heterotic orbifold plus 
these non-perturbative contributions one matches the spectrum of the 
$Z_4^A$  model. The connection with orientifold 
 models was already noticed in ref.~\cite {intri}.  By putting 
together all contributions we are then able to provide a realization of 
the complete perturbative  plus non-perturbative $U(8)^4$  model. Notice 
that we are left with sixteen moduli. The four missing blowing-up modes 
signal the presence of tensor multiplets in a Coulomb phase.

The $Z_6^A$   model is also realized in a similar, but more intricate, 
way. 
Consider the non-modular invariant shift without vector structure given by
\beq
V=\frac1{12}(1,1,1,1,5,5,5,5,3,\cdots,3)
\label{vz6gj}
\eeq
This embedding breaks $SO(32)$ to $U(4)\times U(4) \times U(8)$ and implies
perturbative untwisted hypermultiplets transforming as
$({\bf 6},{\bf 1}, {\bf 1})+({\bf 1},{\bf 6},{\bf 1})+$ 
$({\bf 4},{\bf 1},{\bf 8})+({\bf 1},{\bf 4},{\bf 8}) + 2({\bf 1},{\bf 1},{\bf 1})$.
This matches the (99) sector of $Z_6^A$ orientifold.
To fulfill level-matching in the $\theta^j$ sectors we add an extra vacuum 
energy  $E_B(j)={\frac 14}, \frac13, {\frac 14}$, for $j=1,2,3$.
The mass formula then leads to massless twisted hypermultiplets
transforming as singlets. Specifically, the number of singlet hypermultiplets
in the $\theta^j$ sector is $n_H(j)=2,5,5$, for $j=1,2,3$.

We look next for the non-perturbative piece of the spectrum. The
(55) and (59) content will follow from non-perturbative transitions 
in $\varepsilon_6$ and $\varepsilon_3$ ALE spaces. Recall that in
the $Z_6$ orbifold there is one $\varepsilon_6$, four $\varepsilon_3$
and five $\varepsilon_2$ ALE spaces.
The non-perturbative gauge group comes from small 
instantons at the single $\varepsilon_6$.
Small instantons without vector structure at 
a $Z_6$ singularity generate an enhanced gauge group \cite{intri, bi}  
\beq
U(2\ell_6)\times U(2\ell_6+u_2+u_3-8)\times U(2\ell_6+u_2+2u_3-16)
\label{intriblum2}
\eeq
where $\ell_6$ is an integer large enough to guarantee positive arguments.
Here $u_2$ ($u_3$) is the number of  $\frac3{12}$ ($\frac5{12}$) entries in $V$.
Also, $u_1=(16-u_2-u_3)$ is the number of $\frac1{12}$ entries in $V$.
In this example $u_2=8, u_3=4$.
In addition, there are two tensor multiplets and hypermultiplets
transforming as given in Table~\ref{tablex}.
The $Z_6^A$ orientifold  model is reproduced by setting $\ell_6=2$ at the
$Z_6$ fixed point and $\ell=0$ at the other fixed points.
One then gets a non-perturbative gauge group 
$U(4)\times 
U(8)\times U(4)$ with matter in $4({\bf {4}},{\bf {1}},{\bf {1}})+ 8({\bf 
{1}},{\bf {8}},{\bf {1}})$$ + 4({\bf {1}},{\bf {1}},{\bf {4}}) + ({\bf {4}},
{\bf {8}},{\bf {1}}) +$$ ({\bf {1}},{\bf {8}},{\bf {4}}) + ({\bf {6}},{\bf {1}},
{\bf {1}}) +$$ ({\bf {1}},{\bf {1}},{\bf {6}})$.   
The given multiplicity can now be understood as the gauge quantum 
numbers under the perturbative groups. We also get two tensors from 
this $\varepsilon_6$. The rest of the tensors, as mentioned before, are produced 
after a non-perturbative transition in all four $\varepsilon_3$ ALE's. 
Indeed, these are the $\varepsilon_3$ spaces appearing in the $Z_3^A$ 
 model, which as explained in Section 5, lead to one tensor 
each. 

Notice that the present $Z_6$ orbifold model has 
only fourteen moduli. The six missing moduli signal the presence of six tensor 
multiplets in a Coulomb phase.

We thus see that the heterotic duals of the $Z_M^A$  Type-IIB orientifolds 
can be understood as $Z_M$ orbifolds with non-modular invariant embeddings
without  vector structure in the presence of small instantons.

\subsection{ $Z_4$ Heterotic  Orbifolds with Vector Structure}

Consistent $Z_4$ models with vector structure can also be constructed.
Some examples are listed in Table ~\ref{tablaz4}. 
They correspond to a generic shift
\beq
V=\frac14(1,\cdots, 1, 2, \cdots ,2,0, \cdots , 0)
\label{vz4vs}
\eeq
with $w_1$ $\frac14$ entries and $w_2$ $\frac24$ entries.
The gauge group and corresponding  untwisted matter sector content are
\beqa
G & = & U(w_1) \times SO(2w_2)\times SO(2w_0) \nonumber \\
U & : & ({\bf \ov{w_1}}, {\bf 1}, {\bf 2 w_0}) + 
({\bf w_1}, {\bf 2 w_2},{\bf 1}) + 2({\bf 1}, {\bf 1}, {\bf 1})
\label{z4vs}
\eeqa
where $w_0=16-w_1-w_2$.

Modular invariant models satisfy the constraint
\beq
4w_2+w_1= 2 \ {\rm mod} \ 8
\label{z4modin}
\eeq
This also follows from the instanton numbers $I_f(4)$ and $I_f(2)$
at the $Z_4$ and $Z_2$ fixed points. Indeed, from eq.~(\ref{infix}) we find
\beqa
I_f(4) & = & \ell_4 + \frac{3w_1}8 + \frac{w_2}2 \nonumber \\[0.2ex]
I_f(2) & = & \ell_2 + \frac{w_1}4
\label{indi4}
\eeqa
The total instanton number is $k=4I_f(4)+6I_f(2)$.

It is also interesting to determine the number of states at the $Z_M$ fixed points,
denoted $\cn_f(M)$. From eq.~(\ref{dimin}) we find
\beq
\begin{array}{ccccl}
\cn_f(4) & = & {\rm dim}\ \cam_{inst}(4) + 3 & = &
30\ell_4 +\frac {w_1}{2}(w_1 + w_2 -1) +
w_2( \frac{w_1}{2}+ w_2-1) + 3   \nonumber \\
\cn_f(2) & = & {\rm dim}\ \cam_{inst}(2) + 1 & = &
30\ell_2 + \frac {w_1}{2} (w_1 -1) + 1
\end{array}
\label{dim4}
\eeq
where we have included the contribution from the blowing-up modes.
It is instructive to check how these formulae correctly give the number of
hypermultiplets in the twisted sectors of the modular
invariant $Z_4$ orbifolds in  Table ~\ref{tablaz4} (first and third 
examples). 
To this purpose let us now study the twisted states in more detail.
If $R_1$ denotes the whole set of massless hypermultiplets in the  $\theta$
sector, the total contribution is $4R_1$. The factor 4 takes into account
the four $\theta $ fixed points. In the $\theta ^2$ sector there are twelve 
additional $Z_2$ fixed points, completing six $\theta$-invariant pairs. 
The total contribution from the $\theta^2$ sector is $5R_2 + 3R_3$, where
$R_2$ and $R_3$ are subsets of the massless states. Notice that we have divided
by two so that only particles or antiparticles are counted.
Schematically we have the distribution
\beqa
\theta & : &  4 R_1
\nonumber \\ 
\theta^2 & : &  5 R_2 +3 R_3
\label{z4dist} 
\eeqa
In order to compare with the index theorem results (\ref{dim4}), 
the structure (\ref{z4dist}) can be rewritten to 
show the explicit contribution from the four $\varepsilon _4$ and 
six $\varepsilon _2$  ALE spaces as
\beq
 \theta + \theta ^2 \quad : \quad 4 [R_1 + \oh R_2] +  
 6 [\oh R_2 + \oh R_3] 
\label{z4ale} 
\eeq 
Therefore, the $\cn_f(M)$ are given by
 \beqa
\cn_f(4) & = &  
R_1 + \oh R_2   \nonumber \\ 
\cn_f(2)  & =  &
\oh(R_2 + R_3) 
\label{nfs4} 
\eeqa
The $R_i$ found in the orbifold analysis match exactly the index theorem 
results when the appropriate values for $\ell_4$ and $\ell_2$ are taken. 
These are shown on the last column of Table~\ref{tablaz4}. 
These choices also lead, upon  substitution in equation (\ref{indi4}), 
to a total instanton number of twenty four. 

This exercise can also be performed for the non-modular invariant 
orbifolds that
we are to describe in the following. In each case, we find agreement 
between the spectrum predicted by the index theorem and that found in the 
orbifold. 
Recall that whenever there are fixed points in the Coulomb branch, the 
number of states (actually ${\rm dim} \ {\cal M} _{H}$)  is given 
by ${\cal N}_f(M) = {\rm dim} \ {\cal M} _{inst}+M-1-29P$, where $M = 
2P$ (or $M=2P+1$) \cite{intri}.

A similar analysis is possible for the models without vector structure 
of Section~\ref{dualgj}. Just notice that by computing the number of 
hypermultiplets minus the number of vector multiplets from
the results in Table~\ref{tablex}, we obtain 
 \begin{equation}
{\cal N} _f(4) = {\rm dim} \  {\cal M} _{inst}+3 -29
\label{n4nvs}
\end{equation}
where 
\begin{equation}
{\rm dim} \ {\cal M} _{inst}= 30\ell_4 + \frac12 u_2(u_2-1) 
\label{dm4nvs}
\end{equation}
and similarly \cite{intri} 
\begin{equation}
{\cal N}_f(2)=30\ell_2 +1
\label{nf2nvs}
\end{equation}
We also stress that the contribution of the 
total instanton number found from (\ref{indi4}) plus the number of 
five-branes required always adds to twenty four.

In some  cases it is possible to add an $E_B$ energy shift in
twisted sectors and produce  a transition to a non-perturbative model
in the same way it happened for $Z_3$.
Non-modular invariant shifts are also possible.
Examples of the first situation are the first (standard embedding) case 
as well as the  $U(14)\times U(1)^2$
model corresponding to the shift $V=\frac14(2, 1, \cdots ,1,0)$. In both
cases adding $E_B(1)=\frac14$ in the $\theta$ sector kills some of the
perturbative matter but  singlets that can be identified with 
$K3$ moduli survive. Then,
non-perturbative contributions from standard small
instantons are generated with the adequate content to render the model
consistent. The non-perturbative groups are $Sp(8)$ and $Sp(4)$
respectively.

The last two models in  Table~\ref{tablaz4} are examples in which transitions
to Coulomb phases appear.  For instantons with vector 
structure on a $Z_4$ singularity one expects an enhanced gauge
symmetry \cite{intri} 
\beq
Sp(\ell_4)\times U(2\ell_4+w_1+w_2 -8)\times Sp(\ell_4+\frac{w_1}2+w_2-8)
\label{enh4}
\eeq
There are also
two tensor multiplets and the hypermultiplets shown in Table~\ref{tablex}. 
The models have also $Z_2$ singularities and instantons 
with vector structure on them give an enhanced gauge group
\beq
Sp(\ell_2)\times Sp(\ell_2+{\frac {w_1}2} -4)
\label{enh2}
\eeq
In addition, there is one tensor multiplet and the extra 
hypermultiplets displayed in Table~\ref{tablex}.

The $U(8) \times SO(16)$  ($w_1=8, w_2=0$) model  in the table 
is somewhat similar to the $Z_3$ example considered
in Chapter 5. The shift in this case is, however,  
non-modular invariant. Inclusion of $E_B(1)=\frac1{16}$ and 
$E_B(2)=\frac1{4}$
leads to a consistent model if six tensor multiplets are included.
These tensors can be interpreted as originating from small instantons sitting
at each of the six $\varepsilon _2$ ALE spaces. In fact, as eq.(\ref{enh2}) shows, 
$w_1=8$, $\ell_2=0$ are  critical values. 
Notice that it is also  possible to include a higher  energy shift 
$E_B(1)=\frac5{16}$ in the
$\theta $ sector. The model becomes consistent with  four extra tensor multiplets.
It would signal a transition ${\bf{28}} + {\bf{1}} \rightarrow {\rm 
tensor}$ at each
$\varepsilon_4$ ALE space as we have encountered before. Nevertheless, notice that 
this would not correspond to the situation treated in \cite{intri} since, as 
we remarked before, two
tensor multiplets are expected there  at each $\varepsilon_4$, furthermore 
a critical $\ell_2$ value with $w_1=8$ is not possible.

\begin{table}[p]
\renewcommand{\arraystretch}{1.2}
\footnotesize
\begin{center}
\begin{tabular}{|l|c|c|c|}
\hline
Shift/ Sectors &$( w_2 ,w_1)$ & Group   & $(\ell_4,\ell_2, n_B)/
E_{B}$\\
\hline \hline
 $V=\frac14(1,1,0,\cdots ,0)$  &$ (0,2)$ & $U(2)\times SO(28)$ &  $(3 ,1 ,0)$\\
\hline 
$U: (2,28) +2(1,1)$ &  & & \\ $
\theta: 4[(2,28) + 2(1,1)^* +6(1,1)^*] $ & & & 0\\
$ \theta^2 : 5(2,28) + 16[(1,1)^* +(1,1)^*]$ & & &0 \\
\hline\hline
$V=\frac14(1,1,0\cdots ,0)$ &$ (0,2)$ & $U(2)\times SO(28)$ &
$(1 ,1 ,8)$ \\
\hline 
$U: (2,28) +2(1,1)$ &  & & \\
$\theta: 4[(1,1) +3(1,1)^*] $ & &  & $\frac 14$ \\
$ \theta^2 : 5(2,28) + 16[(1,1)^* +(1,1)^*]$ & &  &0 \\
\hline \hline
$V=\frac14(2,1,\cdots ,1,0)$ & $ (1,14)$ & $U(1)\times U(14)\times 
U(1)$ &$(-2 ,-2,0)$ \\
\hline 
$U: 2(14) + 2(14)+ 2(1)$ &  & & \\
$\theta: 4[(14)+({\ov{14}})+ 2(1)^*+ 2(1)^*] $ & &  & 0\\
$ \theta^2 : (5+3)[(14)+({\ov{14}})+2(1)^*+2(1)^*]$ & &  &0 \\
\hline \hline
 $V=\frac14(2,1,\cdots ,1,0) $ &$ (1,14)$ & $U(1)\times U(14)\times U(1)$
&$(-3 ,-2,4)$ \\
\hline 
$U: 2(14) + 2(14)+ 2(1)$ &  & & \\
$\theta: 4[ (1)+ (1))] $ & &  & $\frac14 $\\
$ \theta^2 : (5+3)[(14)+({\ov{14}})+2(1)^*+2(1)^*]$ & &  &0 \\
\hline \hline
 $V=\frac14(1,1,1,1,1,1,1,1,0,\cdots ,0)$ &$ (0,8)$ & $U(8)\times SO(16)$ 
& $( 0,0,0)$ \\
\hline
$U: (8,16) +2(1,1)$ &  & & \\
$\theta: 4[(28,1) +3(1,1)^*] $ & &  & $\frac1{16}$ \\
$ \theta^2 :  6 \ {\rm Tensors} $ & &  & $\frac14$  \\
\hline
\hline
$ V=\frac 14(1,1,1,1,1,1,1,1,0,\cdots ,0)$  &$ (0,8)$ & $U(8)\times SO(16)$ &
$( 0,0,0)$ \\
\hline 
$U: (8,16) +2(1,1)$ &  & & \\
$\theta: 4[2(1,1)^*]+ 4 \ {\rm Tensors} $ & &  &  $\frac5{16} $\\
$ \theta^2 :  6 \ {\rm Tensors} $ & &  & $\frac14$ \\
\hline
\end{tabular}
\end{center}
\caption{Examples of  consistent $Z_4$  $SO(32)$ orbifold
 models with $E_B\not=0$.
The asterisk indicates twisted states involving left-handed oscillators.
Only the perturbative matter and tensors are shown. $n_B$ gives the number of
five-branes in the bulk.}
\label{tablaz4}
\end{table}

Notice that all the examples discussed in this section correspond to
orbifolds of the $Spin(32)/Z_2$ heterotic.  It would be interesting 
to  explore the  equivalent type of models in the $E_8\times E_8$ case,
for which less is known  about the behaviour of instantons at singularities.

\subsection{Heterotic Dual of  DPGM  $Z_2$ Orientifolds}

We now wish to consider a  $Z_2$  orbifold  of
heterotic $SO(32)$ which yields the same spectrum as the
$Z_2$ orientifold constructed by Dabholkar and Park \cite{dabol2}  
and  model C of Gopakumar and Mukhi \cite{gm}.
This  is a $D=6$, $N=1$
model with gauge group $SO(8)^8$, seventeen tensors and four hypermultiplets.
It can be obtained  in terms of F-theory compactified on the standard
$Z_2\times Z_2$ orbifold,  as a compactification of 
M-theory  on $T^5/Z_2\times Z_2$  and as a type IIB orientifold.
Here  we will obtain it  as a heterotic 
 $SO(32)$  $Z_2$ orbifold (with a non-modular
invariant shift).   We will embed the $Z_2$ twist  in terms of a shift $V$ in 
the $\Gamma _{16} $ lattice  supplemented with two discrete Wilson lines
$a_1$ and $a_2$  as follows
\beqa
V\ =\quad  a_1 & =& {\frac 12} (1,1,1,1,1,1,1,1,0,0,0,0,0,0,0,0) \nonumber \\
a_2  &=&  {\frac 12} (0,0,0,0,1,1,1,1,1,1,1,1,0,0,0,0) 
\label{shigm}
\eeqa
The two Wilson lines break the symmetry down to $SO(8)^4$  whereas 
the $V$ shift projects out  all charged multiplets from the untwisted sector.
Only the four untwisted moduli hypermultiplets remain in that sector. Now, the 
sixteen twisted sectors split into four sets of four fixed points each which
are subject to shifts  $V$, $V+a_2$, $V+a_1+a_2$ and $V+a_1$  respectively.
The first three sets of  four fixed points are all similar, the corresponding shift
has  $w=8$, ${\frac 12}$ entries.  Thus $V^2=2$ and there are no massless
hypermultiplets at any of those twelve fixed points. However,  we already 
mentioned that  the value $w=8$ for embeddings with vector structure 
is critical for five-branes sitting at a $Z_2$ singularity \cite{intri}. 
Indeed, one tensor  and a gauge group $Sp(\ell)\times Sp(\ell+\frac{w}2-4)$
appear  (see Table~\ref{tablex}). 
Since in our case $w=8$, we get one tensor 
for each of the twelve fixed points and no enhanced gauge group for $\ell=0$.

The other four fixed  points with shift $V+a_1$ have a different beahavior.
Indeed, this shift is trivial and hence we have $w=0$ for those fixed points.
As remarked in ref.~\cite{bi2}, five-branes at a $Z_2$ singularity
with $w=0$  give transitions to a Coulomb phase with one tensor multiplet
and a gauge group $Sp(\ell)\times SO(2\ell+8)$
(see Table~\ref{tablex}). Thus, in our case, with $\ell=0$ 
at each of the fixed points we have altogether a non-perturbative
group $SO(8)^4$ and four tensor multiplets.
Putting  all the contributions together we get the total content
$SO(8)^8$, seventeen tensor multiplets and four singlet hypermultiplets.
Notice how the 16 twisted sectors are in a Coulomb phase, twelve of them
with $w=8$ yielding only tensors and the other four 
have $w=0$ yielding in addition the required non-perturbative $SO(8)^4$.

A similar construction can be carried out in the $E_8\times 
E_8$ heterotic, as a $Z_2$ orbifold with the same Wilson line structure 
embedded in $E_8\times E_8$ in a symmetrical way. One obtains the same 
untwisted sector, and the remaining tensors and vector multiplets are 
expected to arise from non-perturbative effects due to small $E_8$ 
instantons (possibly at fixed points). Their dynamics, however, has been 
only partially determined \cite{mv88}, so that a complete check of how 
the spectra match is not available at the moment.  
Notice that the presence of a Wilson line breaking either group to 
$SO(16)\times SO(16)$ shows that both heterotic constructions are related 
by T-duality. An interesting issue in this respect would be to understand 
how the different non-perturbative effects are mapped under this 
transformation.

One can also construct a non-perturbative $Z_2$ orbifold yielding the 
spectrum of the 
model of ref.~\cite{dabol3}  (model B in ref.~\cite{gm}). Consider the   
following shift structure
\beqa
V & = & \frac12 (1,1,1,1,1,1,1,1,0,\cdots ,0) \nonumber \\
A & = & \frac14 (1,\cdots \cdots,1)
\eeqa
The untwisted sector contributes the gauge group $U(8)^2$ and
hypermultiplets transforming as $4 ({\bf 1},{\bf 1}) + ({\bf 8},{\bf {\ov
8}}) + ({\bf {\ov 8}},{\bf 8})$. The fixed points feeling the critical shift
$V$ generate 8 tensors without gauge group, and those feeling the shift
$V+A$ give 8 neutral hypermultiplets. Observe that the total instanton
number adds up to 24, and that all anomalies cancel. The spectrum matches
exactly that in \cite{dabol3}, and upon Higgssing reproduces model B
in \cite{gm}.

\section{$D\!=\!4$, $N\!=\!1$ Non-Perturbative Orbifolds and Chirality Changing
Transitions}

\subsection{$D\!=\!4$, $N\!=\!1$ Non-Perturbative Orbifolds}

In principle, the idea  explored  in previous 
chapters  for the $D=6$  case could be extended
to $D=4$, $N=1$.  One  would 
construct heterotic orbifold vacua 
with perturbative and non-perturbative sectors 
in which the perturbative (but non-modular invariant) 
sector  could  be  understood  in terms of  simple standard orbifold  techniques. 
One must also add a non-perturbative piece, but we face the
problem that non-perturbative phenomena in $N=1$, $D=4$ 
theories are poorly understood at the moment.  In $D=6$  we were guided
by the known results of $D=6$ small instanton dynamics, 
but in $D=4$ there are no clear guidelines.

However, we can concentrate on  
certain  restricted classes of $D=4$ orbifolds
in which much of the structure is expected to be inherited from $D=6$. 
In particular, one can consider  $Z_N\times Z_M$ orbifolds  in $D=4$ with
unbroken $N=1$ supersymmetry.  Such type of orbifolds have two general
classes of twisted sectors, those that leave a 2-torus fixed and those 
that only leave fixed points.  The first type of twisted sectors is
essentially 6-dimensional  in nature,  the twist by itself would lead to
an  $N=2$, $D=4$  theory which would correspond to $N=1$, $D=6$ 
upon decompactification of the fixed torus.  For this type of twisted sectors 
we can use our knowledge of  non-perturbative $D=6$, $N=1$  
dynamics and  the results of the previous chapters.  Twisted sectors of
the second type are purely 4-dimensional in nature and we would need 
extra  information about  4-dimensional non-perturbative dynamics.
To circumvent this lack of knowledge, one can  restrict to a particular class of 
$Z_N\times Z_M$  orbifolds  with gauge embeddings  such 
that these purely 4-dimensional twisted sectors are either absent
or else are not expected to modify  substantially
the structure of the model.    
A  first step could be to try to reproduce  known $D=4$, $N=1$
orientifolds.  A simple example is the $Z_3$, $D=4$ orientifold of
ref.~\cite{bianchi} 
but this example is known to have a perturbative
heterotic dual  \cite{bianchi, kakush} .
 The next simplest $D=4$ Type IIB orientifold is 
the $Z_2\times Z_2$  example of Berkooz and Leigh (BL).  
Below we will describe the heterotic dual of the
$Z_2\times Z_2$ BL orientifold.
We  then consider a class of $Z_3\times Z_3$, $D=4$ 
non-perturbative orbifolds in which certain interesting 
non-perturbative chirality-changing transitions occur.

\subsection{The Heterotic Dual of the $Z_2\times Z_2$ Berkooz-Leigh 
$D=4$ Orientifold}

In the BL model the Type IIB string is twisted
with respect to world-sheet parity combined with the standard $Z_2\times Z_2$
action on six compact (toroidal) dimensions. This model has three sets of
five-branes $5_i$, $i=1,2,3$, whose world-volume fills the four uncompactified
dimensions plus the i-th complex plane.  The largest
gauge symmetry arises when thirty two D-five-branes 
(eight dynamical five-branes) of type $5_i$ coincide at the
same fixed point  (fixed with respect to the $Z_2$ action not touching the
i-th complex plane). Each of these sets induces an $Sp(8)$ maximal gauge symmetry.
The open 99-brane sector also gives rise to an $Sp(8)$ and
the total gauge group is
$Sp(8)_9\times Sp(8)_{5_1}$$\times Sp(8)_{5_2}\times Sp(8)_{5_3}$. There are
also  chiral multiplets transforming as 
$3(\underline {{\bf 120},{\bf 1},{\bf 1},{\bf 1}})$$+
( \underline {{\bf 16},{\bf 16},{\bf 1},{\bf 1} })$, where the underlining means
permutation.  In addition to the gravitational sector, 
the closed string spectrum includes $16\times 3 +6$ 
extra singlet chiral multiplets (moduli) plus the axi-dilaton chiral field.

It is easy to  find  a $Z_2\times Z_2$ orbifold of the $SO(32)$ heterotic string
whose invariant (untwisted) sector corresponds to the (99)-open string sector
of the BL model.  In this case both $Z_2$ actions cannot be realized
simultaneously by shifts in the lattice. One can realize one of the $Z_2$ actions
through the shift
\beq
V = \frac14(1, \cdots ,1)
\label{vz4}
\eeq
This same shift realizes the twist in the heterotic version of
a non-perturbative (from the heterotic point of view)  BSGP model, 
as we described in the previous chapter.  
The other $Z_2$ embedding can be realized in the gauge degrees of freedom 
by a permutation $\Pi $ whose action on the 
sixteen bosonic coordinates $F_I$ is
\beq
\Pi \ :\  (F_1,F_2, \cdots ,F_8,F_9, \cdots,F_{16})\ 
\rightarrow \ (F_9,F_{10}, \cdots ,F_{16},F_1,F_2, \cdots ,F_8) 
\label{fauto}
\eeq
Notice that $\Pi V =V$  so that both actions commute as they should. Each of these 
actions separately break the gauge symmetry down to $U(16)$. But  only a common
$Sp(8)$ subgroup is invariant under both. One can also check that,
associated to the three complex planes there are three untwisted
chiral multiplets transforming as the antisymmetric ${\bf 120}$ of $Sp(8)$. 
Thus, this $Z_2\times Z_2$ indeed reproduces the 
(99) open string spectrum of the $Z_2\times Z_2$ BL orientifold.
Neither $V$ nor $\Pi$ verify the usual modular invariance conditions of 
perturbative orbifolds, in particular $2(V^2-\oh)\not= {\rm even}$.
We are then led to add an appropriate shift $E_B=\frac14$ in each of the 
twisted sectors to recover level matching, as we discussed in 
the $D=6$ examples. With such $E_B$, sixteen singlets
per twisted sector would survive. They will correspond to the twisted moduli.
Non-perturbative effects should give rise to the rest of the particles 
in order to match the orientifold spectra. Each of the twisted sectors 
is indeed a $D=6$ orbifold, so one can apply what is known in
six dimensions and then project. In particular, for  all eight five-branes of 
each  i-th sector  coinciding  at  a fixed point one expects
a non-perturbative group $U(16)_i$, $i=1,2,3$. 
The projection with respect to the other $Z_2$ symmetry should break
each of these groups to $Sp(8)_i$.
One also expects the non-perturbative generation of chiral multiplets 
transforming as ${\bf 16}$ under the perturbative $Sp(8)$ giving rise 
to the charged hypermultiplets appearing in the orientifold construction.  
Thus, one concludes that the  heterotic dual of the BL model is the 
$Z_2\times Z_2$ heterotic orbifold just described,  in which the perturbative part 
is computed in the standard way (modulo a shift in the vacuum energy)
and the non-perturbative piece can be understood in terms of known
$D=6$ small instanton physics.

\subsection{ A Class of Non-Perturbative $Z_3\times Z_3$  Orbifold vacua}

We have seen in previous chapters that the simplest non-perturbative 
heterotic orbifolds  in $D=6$  are obtained for $Z_3$ twists. Thus, it is
natural to consider $D=4$  $Z_3\times Z_3$ orbifolds  with gauge embeddings of the 
restricted form described in Chapter 3, with length-squared sufficiently small.
Let us first review a few points about generic
$Z_3\times Z_3$ orbifolds  (see ref.~\cite{fiq} for more details). 
The point group is generated by twists $\theta $ and $\omega$ with twist vectors
given by $a=\frac13(0,1,0,-1)$ and $b=\frac13(0,0,1,-1)$
respectively. The gauge embedding shifts are given by 16-dimensional
vectors $A,B$ satisfying  $3A,3B\in \Gamma _{16}$. Apart from the 
untwisted sector there are eight twisted sectors. Those with gauge embeddings  
$A,B$, $(A-B)$ and their inverses are 6-dimensional in nature.
The remaining two twisted sectors have twist vector $(a+b)=\frac13(0,1,1,-2)$
and gauge shift $(A+B)$ and their inverses. They comprise a
standard $Z$ orbifold  with twenty seven fixed points. 
These sectors are purely 4-dimensional in nature, and we want to circumvent them.
In  some particular cases, as we will show below, they have no massless 
fields and hence we can proceed by using essentially only $D=6$ information. 
In most of the cases, however, there are massless fields. In this case
we should  restrict to models with shifts $A,B$ verifying the usual
$Z$-orbifold modular invariance constraint for the $(A+B)$ shift, namely
In this way we will avoid purely 4-dimensional non-modular invariant 
shifts.  On the other hand, the $A,B$ and $(A-B)$ shifts
can be allowed to violate modular invariance constraints.
We now present several specific examples with a variety of interesting properties.

\smallskip\noindent
{\it Example 1}
\smallskip

\noindent
This model is interesting because it shows the existence of non-perturbative 
$D=4$ transitions between perturbative and non-perturbative  vacua
in which the particle spectrum changes. It also seems to have a  Type IIB
orientifold dual.
Consider the shifts 
\beqa
A & = & \frac13(1,1,1,1,1,1,1,1,0,0,0,0,0,0,0,0) \nonumber \\ 
B & = & \frac13(0,0,0,0,1,1,1,1,1,1,1,1,0,0,0,0)
\label{abex1}
\eeqa
in $\Gamma_{16}$. Notice that each of these shifts by itself would give rise in 
$D=6$ to  the  $U(8)\times SO(16)$ models discussed in the Section 5 which
are in turn related to the $Z_3^A$ GJ orientifold. Thus, this
example is expected to be dual to a $D=4$ $Z_3\times Z_3$ Type IIB orientifold
generalization of the $Z_3^A$, very much like the $Z_2\times Z_2$ BL orientifold
is a $D=4$ generalization of the BSGP model (as we were finishing up this paper 
ref.~\cite{kakush2} appeared in which the Type-I dual of this model is 
explicitly constructed).
One can easily check that for these specific $A$ and $B$ the usual modular 
invariance constraints are verified in all twisted sectors. 
The gauge group is $U(4)^3\times SO(16)$ and 
the complete chiral multiplet spectrum (except for the dilaton $S$
and untwisted moduli) is displayed in Table~\ref{tabla8}. 

\begin{table}[htb]
\renewcommand{\arraystretch}{1.2}
\begin{center}
\begin{tabular}{|c|c|}
\hline
Sector & Representation   \\
\hline\hline
$U_1$ & (4,1,1,$8_v$)+(6,1,1,1) +(1,$\overline 4$, $\overline 4$,1)  \\
\hline
$U_2$ & (1,4,1,$8_v$)+(1,6,1,1)+($\overline 4$, 1, $\overline 4$,1)  \\
\hline
$U_3$ & (1,1,4,$8_v$) + (1,1,6,1)+ ($\overline 4 $,$\overline 4$, 1,1)  \\
\hline
$A,\bar A$ &  9(6,1,1,1)+9(1,6,1,1)+18(1,1,1,1)  \\
\hline
$B,\bar B$ &  9(1,6,1,1)+9(1,1,6,1) + 18(1,1,1,1)  \\
\hline
$A-B,{\bar A}-{\bar B}$ &  9(6,1,1,1)+9(1,1,6,1) + 18(1,1,1,1)  \\
\hline
$A+B$ &  27(1,1,1,1) + 27(1,1,1,$8_s$) \\
\hline
\end{tabular}
\end{center}
\caption{Spectrum of  the $Z_3\times Z_3$, $U(4)\times U(4)\times U(4)\times SO(8)$ 
perturbative model.}
\label{tabla8}
\end{table}

But we can equally consider a non-perturbative orbifold in which the $D=6$ 
subsectors $A$, $B$ and $A-B$ have a  left-handed vacuum energy shifted by 
$\frac13$. 
As we showed in Chapter 5, this corresponds to a non-perturbative $D=6$ 
vacuum in which twisted sectors have just one singlet and  nine tensor
multiplets appear.  Non-perturbative $D=6$ transitions exist between the
modular invariant  $E_B=0$ and  the $E_B=\frac13$ models.  Thus, 
the corresponding $D=4$ non-perturbative orbifold would just have 
some singlets in the twisted $A,B$,$A-B$  sectors and there should be
non-perturbative transitions in which all the states transforming as ${\bf 6}$-plets
in those twisted sectors disappear from the massless spectrum.  
The untwisted and $A+B$-twisted sectors will remain unchanged.
This would just be a 4-dimensional version of the $D=6$ transitions in 
which twenty nine hypermultiplets are converted into one tensor multiplet.
Notice how the non-Abelian $SU(4)^3\times SO(8)$ anomalies cancel since the
untwisted sector is (non-trivially) anomaly-free by itself. The existence of $U(1)$
anomalies is  expected,  but notice that now a generalized 4-dimensional 
Green-Schwarz mechanism can take place since there will be extra 
singlet fields (associated to the $9\times 3$ tensors of the $D=6$ twisted sectors)
which will couple to the gauge groups in a non-universal manner.

Let us finally remark that no special non-perturbative effects are expected to
arise from the $A+B$ sector. Indeed this sector is identical to the twisted sector
of a $Z_3$, $D=4$ orbifold considered  in ref.~\cite{bianchi, kakush}.
These authors showed this orbifold to be dual to a certain Type IIB $Z_3$, 
$D=4$ orientifold which has the relevant characteristic of having no 
five-branes (very much like the $Z_3^A$ GJ model).  This implies that one does not 
expect new non-perturbative  charged hypermultiplets nor enhanced gauge 
group coming from this sector.

\smallskip\noindent
{\it Example 2}
\smallskip

\noindent
This second example illustrates how there can be
non-perturbative transitions in which chiral  generations disappear
from the massless spectrum. In this specific case the final non-perturbative
$D=4$ model will have three $E_6$ generations.
Consider the  $Z_3\times Z_3$ orbifold on $E_8\times E_8$ with gauge shifts 
\beqa
A & = & \frac13(1,1,0, \cdots ,0)\times (0, \cdots, 0) \nonumber \\ 
B & = & \frac13(0,1,1,0, \cdots, 0)\times (0, \cdots ,0)
\label{abex2}
\eeqa
This leads to a perfectly modular invariant orbifold with gauge group 
$E_6\times U(1)^2\times E_8$.
However, we are going to consider the particular version of this orbifold 
with discrete torsion first considered in ref.~\cite{fiq}.  This model has the 
particular property that there is no $(A+B)$  massless twisted sector, 
all particles are projected out.  In this way we get rid of the sector which is
purely 4-dimensional. The model has now three ${\bf 27}$'s in the untwisted 
sector and nine ${\bf \ov{27}}$'s in each of the sectors $A,B$ and $A-B$. 
Hence, altogether the model has twenty four
net antigenerations. Very much like in the previous 
example, we can now consider a  non-perturbative orbifold in which the 
$D=6$ subsectors $A,B$ and $A-B$ have a left-handed vacuum energy shifted by 
$\frac13$. As we showed in Chapter 3, this corresponds to a non-perturbative 
$D=6$ vacuum with just singlets in the twisted sectors and eighteen five-branes 
(leading to tensor multiplets) in each of the three twisted sectors. 
Therefore, all the twenty seven antigenerations of the twisted sectors 
disappear from the spectrum and we are only left  with three
$E_6$ generations coming from the untwisted sector, plus singlets. 
As in the previous example the $U(1)$'s will now be 
anomalous but there will be extra chiral singlets, coming from
the tensors, with non-universal couplings to the gauge fields which will lead to 
a generalized version of the GS mechanism in $D=4$.

This example shows that the number of chiral generations is not invariant under 
non-perturbative effects. Vacua with different number of generations are
connected. 
 
\smallskip\noindent
{\it Example 3}
\smallskip

\noindent
We also expect chirality changing transitions 
in compactifications of heterotic $SO(32)$ theory.  Consider the 
$Z_3\times Z_3$ orbifold with gauge shifts 
\beqa
A & = & \frac13(1,1,1,1,1,-2,2,-1,0, \cdots ,0) \nonumber \\
B & = & \frac13(0,0,0,0,0,0,1,1,0,\cdots ,0)
\label{abex3}
\eeqa
This leads to a modular invariant $D=4$ model
with gauge group $SO(16)\times SU(6)\times SU(2)\times U(1)^2$ and 
chiral multiplet spectrum  shown in Table~\ref{tabla9}.

\begin{table}[htb]
\renewcommand{\arraystretch}{1.2}
\begin{center}
\begin{tabular}{|c|c|}
\hline
Sector & Representation  \\
\hline\hline
$U_1$ & (16,6,1) + (1,$\overline{15}$,1)  \\
\hline
$U_2$ & (1,6,2)  \\
\hline
$U_3$ & (1,6,2)   \\
\hline
$A,\bar A$ & 9(1,6,2)+ 9 (1,$\overline{15}$,1)  + 18(1,1,1)  \\
\hline
$A-B,{\bar A}-{\bar B}$ &9 (1,6,2) + 9(1,$\overline{15}$,1) + 18(1,1,1)  \\
\hline
$B,\bar B$ & 18 (1,6,2)  + 54 (1,1,1) \\
\hline
$A+B$ & 27  (1,$\overline{15}$,1) +27 (1,1,1)  \\
\hline
\end{tabular}
\end{center}
\caption{Spectrum of $Z_3\times Z_3$, $SO(16)\times SU(6)\times SU(2)$
 model.}
\label{tabla9}
\end{table}

Now, the $A$ and $A-B$  shifts by themselves would have given rise to the
$U(8)\times SO(16)$  model in $D=6$. We know that that model  has transitions
to a model with  vacuum shift $E_B=\frac13$ in which only singlets 
(and  tensor multiplets) appear in the  twisted sector.  Thus, one would expect
the existence of  $D=4$ non-perturbative transitions in which  the 
$SU(6)$ chiral generations appearing in the  $A$ and $A-B$ twisted  sectors
disappear  from the massless spectrum. This would be analogous to the
transitions discussed in the previous example.  Unfortunately, in the present example 
we  do not know  if extra non-perturbative effects associated to the
twisted $A+B$ sector exist or not,  but it seems reasonable to expect the
existence of these chirality changing transitions  coming from the 
$A$ and $A-B$ sectors.

In conclusion, our lack of a better
knowledge of non-perturbative  dynamics in $D=4$, $N=1$
does not allow us to make a straightforward generalization to this case. 
However, there are certain classes of $Z_N\times Z_M$ $D=4$
non-perturbative orbifolds in which  interesting conclusions
can be obtained on the basis of  $D=6$ information. Among the most relevant ones
is the observation that  there are non-perturbative transitions  which change the
number of chiral generations.

\section{Final Comments and outlook}

In this paper we have studied  a class of  non-perturbative $D=6$, $N=1$ orbifolds 
of $Spin(32)/Z_2$ and $E_8\times E_8$ heterotic strings.  They are obtained 
by modding the gauge degrees of freedom of these theories by $Z_M$
actions not obeying the usual (perturbative) modular invariance
constraints.  These models have  perturbative and  non-perturbative pieces.
The massless spectrum of the perturbative sector is obtained by the  usual
string mass formula, with  the  $Z_M$ twisted sector vacua subject to  a shift 
in the vacuum energy.  This is due to the presence of a non-vanishing 
flux of the antisymmetric field $H$ at the fixed points.
The number  of hypermultiplets found in  this way 
matches with the 
index theorem formulae for instantons on  $Z_M$ ALE spaces.
The non-perturbative sector is obtained from the knowledge of  small
instanton dynamics.  Some of the models are orbifold equivalents of
 the vacua obtained from smooth $K3$ compactifications in the presence of
small instantons.  For those orbifold models the non-perturbative sector is
provided by the world-volume theories  of  wandering 
five-branes with the world-volume
in the uncompactified dimensions.  In some other cases the non-perturbative sector 
includes  five-branes which are stuck  at the $Z_M$ fixed points giving more
exotic physics including tensor multiplets  as described in ref.~\cite{intri}.
The heterotic duals of the $Z_M^A$ Type IIB orientifolds can be understood 
in this way but there are many other models which can be constructed.

We think that the present class of non-perturbative 
vacua  are  of  practical interest not only because
it provides the heterotic duals of a number of models 
but also because of its simplicity. One just uses
familiar orbifold techniques  supplemented by information about 
small instanton physics.  Many models can be obtained for the
different $Z_M$'s  by using  the diverse possible embeddings in the 
gauge degrees of 
freedom.  In contrast, notice that in Type IIB orientifolds the gauge embedding
(up to the addition of Wilson lines and/or discrete torsion)
is essentially uniquely fixed by tadpole cancelation.
Also, a technique like F-theory is by far more general and powerful  but 
is slightly cumbersome if one is interested in knowing in detail the 
hypermultiplet 
spectra and the charges under the gauge groups, especially $U(1)$'s, for 
particular points of F-theory moduli space. 

As we said above, the non-perturbative sector of these orbifold models is 
obtained from  small instanton dynamics. In  orbifolds  like these, 
information about   
five-branes wandering in the bulk and five-branes stuck to the 
fixed points of the orbifold are both needed.  Our knowledge about the 
latter is only partial.  When a large enough  number  of point-like 
instantons are piled on a
$Z_M$  singularity   in a $Spin(32)/Z_2$   compactification,
there is a Coulomb phase in which a known spectrum of
tensor, gauge and hypermultiplets appear.  Such complete information
(particularly for the hypermultiplet sector)  is not yet available for the
$E_8\times E_8$ case.   Furthermore  we do not know, 
except for some cases, the expected 
non-perturbative spectrum when the number of small instantons on the
singularity is smaller than the critical value.  It is reasonable to expect that 
this information will soon become available and a larger class of 
non-perturbative orbifolds along the lines considered in this paper 
will be constructed.

In addition to the previous comment, there are several interesting questions 
to pose. For example, what happens if  the perturbative sector of our model 
is not a simple geometrical orbifold  but  a  (non-modular invariant) 
asymmetric orbifold or, in general, an arbitrary CFT. Most likely, there 
are similar non-perturbative effects that render the theory consistent, 
and it would be interesting to determine their nature.
Another interesting point is the extension of these ideas
to $D=4$, $N=1$ vacua, which is of more direct phenomenological 
interest.  We have seen that in certain  $N=1$, $D=4$,  $Z_N\times Z_M$  
 orbifolds  one can  obtain some non-perturbative vacua  by  using
$D=6$ information. One finds that  the existence of $D=6$ transitions in which 
one tensor multiplet transmutes into  twenty nine charged hypermultiplets, 
implies in $D=4$ the existence  of transitions in which chiral generations
({\it e.g.}  {\bf 27}'s of $E_6$ or 
 {\bf 15}+ 2 $\cdot   {\overline {\bf 6}})$ of $SU(6)$)  
change into singlets.
Thus  $D=4$, $N=1$ vacua with different number of chiral  generations are
non-perturbatively connected.  Of course, this is of   relevance if
one is eventually interested in describing the observed physics in terms 
of string (or M-theory)  dynamics.

\vskip0.6cm

\centerline{\bf Acknowledgments}
\bigskip

We thank  F. Quevedo for useful discussions.
G.A. thanks   ICTP and the Departamento de F\'{\i}sica Te\'orica
at UAM for hospitality and financial support.
A.M.U, L.E.I. and G.V. thank CICYT  (Spain) for financial support.
A.F. thanks CONICIT (Venezuela) for a research grant S1-2700.
A.M.U. thanks the Governement of the Basque Country for financial support.


\begin{thebibliography}{99}
%
%
\bibitem{witsm}
E.~Witten, \NPB{460} {96} {541}, hep-th/9511030.
%
\bibitem{dmw}
M.~Duff, R.~Minasian and E.~Witten, \NPB{465} {96} {413}, hep-th/9601036.
%
\bibitem{dhvw}
L.~Dixon, J.A.~Harvey, C.~Vafa and E.~Witten, \NPB{274} {86} {285}.
%
\bibitem{orbi}
L.E.~Ib\'a\~nez, J.~Mas, H.P.~Nilles and F.~Quevedo,
\NPB{301}{88}{157};\\
A.~Font, L.E.~Ib\'a\~nez, F.~Quevedo and A.~Sierra,
\NPB{331}{90}{421}.
%
\bibitem{intri}
K.~Intriligator, hep-th/9702038.
%
\bibitem{gj}
E.~Gimon and C.~Johnson,
\NPB{477}{96}{715}, hep-th/9604129.
%
\bibitem{dabol1}
A.~Dabholkar and J.~Park, \NPB{477}{96}{701}, hep-th/9604178.
%
\bibitem{sw6d}
N.~Seiberg and E.~Witten,
\NPB {471} {96} {121}, hep-th/9603003.
%
\bibitem{mv1}
D.~Morrison  and C.~Vafa, \NPB{473} {96} {74}, hep-th/9602114. 
%
%
\bibitem{kasi}
S.~Kachru and E.~Silverstein, hep-th/9704185.
%
\bibitem{dabol2}
A.~Dabholkar and J.~Park, \PLB{394}{97}{302}, hep-th/9607041. 
%
\bibitem{gm}
R.~Gopakumar and S.~Mukhi, \NPB{479}{96}{260}, hep-th/9607057.
%
\bibitem{bso}
M.~Bianchi and A.~Sagnotti,   \NPB{361} {91} {519}.
%
\bibitem{gp}
E.~Gimon and J.~Polchinski,
Phys.Rev. D54 (1996) 1667, hep-th/9601038. 
%
\bibitem{bl}
M.~Berkooz and R.~G.~Leigh, \NPB{483} {97} {187}, hep-th/9605049.
%
\bibitem{walton}
M.~A.~Walton, \PRD{37} {87} {377}. 
%
\bibitem{erler}
J.~Erler, J. Math. Phys. 35 (1994), 1819, hep-th/9304104.
%
\bibitem{afiq}
 G.~Aldazabal, A.~Font, L.~E.~Ib\'a\~nez and F.~Quevedo,
\NPB{461} {96} {85}, hep-th/9510093.
%
\bibitem{kv}
S.~Kachru  and C.~Vafa, \NPB{450} {95} {69}, hep-th/9505105. 
%
\bibitem{afiu}
 G.~Aldazabal, A.~Font, L.~E.~Ib\'a\~nez and A.~Uranga,
hep-th/9607121.
%
\bibitem{berkooz}
M.~Berkooz, R.~G.~Leigh, J.~Polchinski, J.~H.~Schwarz, N.~Seiberg and  E.~Witten,
\NPB{475} {96} {115}, hep-th/9605184.
%
\bibitem{aspfz2}
P.~Aspinwall, hep-th/9612108.
%
\bibitem{gsw}
M.~B.~Green, J.~H.~Schwarz and P.~C.~West, \NPB {254} {85} {327}.
%
\bibitem{dasmu}
K. Dasgupta and S. Mukhi, \NPB{465}{96}{399}, hep-th/9512196.
%
\bibitem{bz}
J.~Blum and A.~Zaffaroni, \PLB{387}{96}{71}, hep-th/9607019. 
%
\bibitem{blum}
J.~Blum, \NPB{486}{97}{34}, hep-th/9608053.
%
\bibitem{senft}
A.~Sen, \NPB{489}{97}{139}, hep-th/9611186; 
\PRD{55}{97}{7345}, hep-th/9702165.
%
\bibitem{hor} 
A. Sagnotti, in Cargese 87, 
{\it Strings  on Orbifolds}, 
ed. G. Mack et al. (Pergamon Press, 1988) p. 521;\\
P.~Horava, \NPB{327} {89} {461}; \PLB{231} {89} {251};\\
J.~Dai, R.~Leigh and J.~Polchinski, Mod.Phys.Lett. A4 (1989) 2073;\\
R.~Leigh, Mod.Phys.Lett. A4 (1989) 2767;\\
J.~Polchinski, \PRD{50} {94} {6041}, hep-th/940703.
%
\bibitem{bs} G.~Pradisi and A.~Sagnotti, \PLB{216} {89} {59};\\
M.~Bianchi and A.~Sagnotti, \PLB{247} {90} {517}.
%
\bibitem{bianchi}
C.~Angelantonj, M.~Bianchi, G.~Pradisi, A.~Sagnotti and Ya.S.~Stanev,
\PLB{385} {96} {96}, hep-th/9606169.
%
%
\bibitem{kakush}
Z.~Kakushadze, hep-th/9704059.
%
\bibitem{sagn}
A.~Sagnotti, \PLB{294} {92} {196}. 
%
\bibitem{quivers}
M.~Douglas and G.~Moore, hep-th/9603167.
%
\bibitem{bi}
J.~Blum and K.~Intriligator, hep-th/9705030.
%
\bibitem{bi2}
J.~Blum and K.~Intriligator,  hep-th/9705044.
%
\bibitem{mv88}
 P.~Aspinwall and D.~Morrison, hep-th/9705104. 
%
\bibitem{canpera}
P.~Candelas, E.~Perevalov and G.~Rajesh, hep-th/9704097.
%
\bibitem{mirror}
E.~Perevalov and G.~Rajesh, hep-th/9706005.
%
\bibitem{aps}
M.~F.~Atiyah, V.~K.~Patodi and I.~M.~Singer, Math. Proc. Camb. Phil. 
Soc.~77 (1975) 43; 77 (1975) 405; 79 (1976) 1. 
%
\bibitem{roma}
M.~Bianchi, F.~Fucito, M.~Martellini and G.~Rossi, 
\NPB{473} {96} {367}, hep-th/9601162.
%
\bibitem{gh}
O.~Ganor and A.~Hanany, \NPB{474} {96} {122}, hep-th/9602120.
%
\bibitem{dabol3}
A.~Dabholkar and J.~Park, \NPB{472}{96}{207}, hep-th/9602030.
%
\bibitem{kakush2}
Z. Kakushadze and G. Shiu, hep-th/9706051.
%
\bibitem{fiq} 
 A.~Font, L~E.~Ib\'a\~nez and F.~Quevedo.
\PLB {217}{89}{272}.
%








\end{thebibliography}
\end{document}